\documentclass[11pt, showpacs]{article}
\pdfoutput=1
\usepackage{amsmath,amssymb,amsfonts,bm,marginnote,cite,graphicx}
\usepackage[colorlinks=true, pdfstartview=FitV, linkcolor=black, citecolor=black, urlcolor=black]{hyperref}
\usepackage[usenames,dvipsnames]{xcolor}

\usepackage{cite}
\usepackage{graphicx}
\usepackage{epstopdf}
\usepackage[justification=centering]{caption}
\usepackage{subcaption}
\usepackage{dcolumn}
\usepackage{bm}
\usepackage{bbm}
\usepackage{amsmath}
\usepackage{amscd}
\usepackage{amssymb}
\usepackage{mathrsfs}
\usepackage{dsfont}
\usepackage{comment}
\usepackage{setspace}
\usepackage[margin=0.75in]{geometry}
\usepackage[colorlinks=true,linkcolor=black,citecolor=black]{hyperref}
\usepackage{slashed}
\usepackage{tensor}
\usepackage{mathrsfs}
\usepackage{cancel}
\usepackage{hyperref}
\usepackage{empheq}
\usepackage{youngtab}
\usepackage{empheq}
\usepackage{float}
\usepackage[all]{xy}

\edef\marginnotetextwidth{\the\textwidth}

\newcommand{\thistitle}{
	Unitary Networks from the Exact Renormalization\\ of Wave Functionals 
	}

\newcommand{\addressuiuc}{
	Department of Physics, University of Illinois,
 	1110 West Green St., Urbana IL 61801, U.S.A.
	}

\newcommand{\disent}{disentangler}

\newcommand{\be}{\begin{equation}}
\newcommand{\ee}{\end{equation}}
\newcommand{\beq}{\begin{eqnarray}}
\newcommand{\eeq}{\end{eqnarray}}
\newcommand{\bea}{\begin{eqnarray}}
\newcommand{\eea}{\end{eqnarray}}
\newcommand{\beqn}{\begin{eqnarray}}
\newcommand{\eeqn}{\end{eqnarray}}

\def\pa{\partial}

\newcommand*\widefbox[1]{\fbox{\hspace{2em}#1\hspace{2em}}}

\def\Tr{\text{Tr}}
\newcommand{\hcO}{\hat{\mathcal{O}}}
\newcommand{\vx}{\vec{x}}
\newcommand{\circp}{\circ}
\newcommand{\pdot}{\cdot}
\newcommand{\dis}{\boldsymbol{K}}
\newcommand{\dil}{\boldsymbol{L}}

\begin{document}

\title{\thistitle}
\author{
	{Jackson R. Fliss$^1$, Robert G. Leigh$^1$ and Onkar Parrikar$^{1,2}$}\\
	\\
	{$^1$\small \emph{\addressuiuc}}
	\\
	{$^{2}$\small \emph{David Rittenhouse Laboratory, University of Pennsylvania,
	209 S. 33rd Street, Philadelphia PA 19104, U.S.A.}}
\\}
\date{\today}
\maketitle\thispagestyle{empty}

\begin{abstract}
The exact renormalization group (ERG) for $O(N)$ vector models (at large $N$) on flat Euclidean space can be interpreted as the bulk dynamics corresponding to a holographically dual higher spin gauge theory on $AdS_{d+1}$. This was established in the sense that at large $N$ the generating functional of correlation functions of single trace operators is reproduced by the on-shell action of the bulk higher spin theory, which is most simply presented in a first-order (phase space) formalism. In this paper, we extend the ERG formalism to the wave functionals of arbitrary states of the $O(N)$ vector model at the free fixed point. We find that the ERG flow of the ground state and a specific class of excited states is implemented by the action of unitary operators which can be chosen to be local. Consequently, the ERG equations provide a continuum notion of a tensor network. We compare this tensor network with the entanglement renormalization networks, MERA, and its continuum version, cMERA, which have appeared recently in holographic contexts.  In particular the ERG tensor network appears to share the general structure of cMERA but differs in important ways.  We comment on possible holographic implications.
\end{abstract}

\newpage
\section{Introduction}

Holography is a correspondence between apparently very different physical systems. A fruitful way to think of this correspondence is that it encodes the renormalization group (RG) flow of a (generally strongly coupled) quantum field theory into a geometric system, generally in higher dimension, that under certain circumstances may be classical or semi-classical. Conformal fixed points of the field theory's RG flow possessing a conformal group $G$ correspond to systems on geometries whose isometry coincides with $G$, at least locally. Although first introduced as a duality between strongly coupled large $N$ field theories and classical gravitational theories with matter, it is known that the correspondence runs more deep. In fact, free $O(N)$ vector models truncated to the single trace sector are now known to be dual to semi-classical theories with an enormous gauge symmetry, generally known as higher spin theories \cite{Klebanov:2002ja, Sezgin:2002rt, Leigh:2003gk, Giombi:2009wh, Giombi:2012ms}. In Refs. \cite{Leigh:2014tza,Leigh:2014qca}, it was argued that the gauge symmetry of the higher spin theories has an origin in a bi-local, linear symmetry of free field theories, and employing the Polchinski exact renormalization group (ERG) methods, a dual canonical bulk theory was derived. 
In this description, the bulk fields correspond to a connection 1-form for the bi-local symmetry plus an adjoint-valued 0-form. There is a solution of the bulk theory in which the adjoint 0-form field vanishes and the connection is taken to be flat; this solution corresponds to the conformal fixed point, and the flat connection encodes the geometry associated with that fixed point (e.g., $AdS_{d+1}$ when the fixed point has relativistic $d$-dimensional symmetry).

This holographic description is in fact effective: many of the usual elements of the AdS/CFT dictionary can be seen to emerge from the above construction. In particular, there is an exact action for the bulk higher spin theory\footnote{More precisely, the bulk dynamics is described in the Hamiltonian formalism, which can then be rewritten in terms of a phase space action.} and it has been shown explicitly that the action when evaluated on-shell precisely generates all of the correlation functions of the field theory. The `holographic dictionary' of this higher spin theory contains many familiar entries, along with several features (such as the structure of `Witten diagrams') which are somewhat different, but appropriate to the first order connection formalism. Furthermore, it has been shown \cite{Jin:2015aba} that at the linearized order, the bulk equations of motion in this higher spin theory are canonically equivalent to the Fronsdal equations, which is a basic sanity check from the point of view of group theory. 

The story thus far has focussed on the generating functional of correlators in the vacuum state, which the original GKPW dictionary was built around. The goal of the present paper is to extend the ERG analysis to the renormalization group flows of the \emph{wave functionals} of the vacuum and a class of excited states; see  \cite{Symanzik:1981wd, Luscher:1985iu, Cooper:1987pt, D.Minic:1996aa} for previous work along these lines. For simplicity, we will confine ourselves in this paper to the $O(N)$-singlet sector of the bosonic vector model, with the excited states created by the action of single-trace operators acting on the vacuum. The main result which will emerge from this analysis is a formulation of ERG for states in terms of a continuum analogue of a \emph{tensor network}. For instance, the ERG flow equation for the vacuum state takes the form 
\beq 
z\frac{\pa}{\pa z}\Big|\Omega(z)\Big\rangle =  i\Big(\dis(z) + \dil(z)\Big) \Big|\Omega(z)\Big\rangle,
\eeq
where both $\dis$ and $\dil$ are local (in position space) unitary operators; a similar equation works out for excited states as well. This unitary action can be naturally extended to $O(N)$-singlet excited states by requiring as an RG principle that these states, as well, have unitary flows.  This requirement is, in fact, equivalent to requiring that the sources preparing the states flow according to their standard beta functions. The action of the operator $\dis$ on the state has the effect of freezing out the high-energy modes while leaving the low-energy modes untouched. For excited states, this has an interesting interpretation in terms of entanglement -- the operator $\dis$ removes all entanglement between high-energy and low-energy modes. In other words, $\dis$ acts as a disentangler in \emph{momentum space}. This is to be contrasted with the usual description of tensor networks such as MERA \cite{Vidal:2007hda, Pfeifer:2008jt} or cMERA \cite{Haegeman:2011uy, Nozaki:2012zj}, where the central theme is to disentangle the state in position space. 

As mentioned previously, the field theory of interest in this paper, namely the single-trace sector of the $d$-dimensional $O(N)$ vector model, is holographically dual to a higher spin theory on $AdS_{d+1}$. Indeed, there is by now a sizable literature claiming a deep relation between tensor networks and holography; see \cite{Swingle:2009bg, Evenbly:2011, Swingle:2012wq, Nozaki:2012zj, Mollabashi:2013lya, Lee:2015vla, Pastawski:2015qua, Czech:2015kbp} and references there-in. It is therefore clearly an interesting question whether we can shed light on the $AdS$/tensor network correspondence within our framework. However the spirit of the present paper is primarily field theoretic -- making contact with the holographic descriptions presented previously in \cite{Leigh:2014tza,Leigh:2014qca} will be left to future work.  Although we are not discussing real-space entanglement renormalization in the present work, we should perhaps remark in passing that for theories with higher spin duals, the holographic dictionary to compute real-space entanglement (for excited states) is not known in general dimension. Furthermore, these dual higher spin theories are not geometric in the conventional sense, and so it is unclear to what extent one can recover conventional geometries from the tensor network constructions for free theories.

The rest of the paper is organized as follows. Because we want to employ {\it exact} renormalization methods, in Section 2 we recall the construction of wave functionals in continuum free field theories. Although much of this discussion is standard material, we provide it here for completeness. In particular, we consider $N$ scalar fields and define the generator of such states by introducing sources for all $O(N)$-singlet single trace operators. As well, we give a short review of the implementation of ERG for the partition function, regarded as a functional of the operator sources. In Section 3, we discuss the implementation of ERG in the context of wave functionals and transition amplitudes. In Sections 3.1 and 3.2 the renormalization scale dependence of the ground state and excited states, respectively, is determined. The form of these equations suggest an interpretation in terms of a scale transformation and a {\it disentangling} operation (in momentum space). In Section 4, we discuss this interpretation of the solutions of the RG equations and make comparisons with Wilsonian RG and entanglement RG. We have also included a number of Appendices, which contain background material on and further details on the calculation of wave functionals in field theories, on the form of suitable regulators and further details of the exact renormalization group formalism.

\section{$O(N)$ Singlet wave functionals in Vector Models}\label{SingWFsect}
We consider a relativistic scalar field theory with $N$ scalars $\phi^a(t,\vec x)$ (where $a = 1, \cdots N$) on $d$ dimensional Minkowski spacetime ($D=d-1$ will denote the number of spatial dimensions). To construct a basis for the Hilbert space, we introduce a foliation in terms of space-like hypersurfaces, $\{\Sigma_t\}$ (where $t$ is the Minkowski time). On a particular hypersurface, $\Sigma$, we introduce a basis $|\varphi^a(\vec x)\rangle$ for the Hilbert space, with $\varphi^a(\vec x)\in L^2(\Sigma)$. 

For any generic state, it is useful to work with its overlap with the basis states $|\varphi^a(\vec x)\rangle$, that is,  the wave functional corresponding to the state.  For the theory under consideration, there is a simple and explicit construction of the wave functionals corresponding to the ground state and a class of $O(N)$-singlet excited states, in terms of a path integral with fixed boundary conditions on $\Sigma$; we will review this construction below. The action is given by
\begin{equation} \label{Fullaction}
S_\phi=\frac{1}{2}\sum_{a=1}^N\int_M dt\,d^{D}\vec x\,\phi^a(t,\vec x)\,\Box\phi^a(t,\vec x)+\frac{1}{2}\sum_{a=1}^N\int_{\Sigma}d^{D}\vec x\ \phi^a(\vec x)\,n^{\mu}\pa_{\mu}\phi^a(\vec x).
\end{equation}
where we are using the mostly plus Lorentzian signature: $\Box\equiv \pa_{\mu}\pa^{\mu}=-\pa_t^2+\vec\nabla^2$.  From here on we will drop the $O(N)$ indices and leave their sum implicit.\\
\\
From a classical point of view, the boundary term in \eqref{Fullaction} is necessary to obtain a well-defined variational principle, consistent with fixing $\phi\Big|_{\Sigma}=\varphi$, while in the path integral language, these terms enforce appropriate boundary conditions. 
Since we are in free field theory we have the luxury of performing the path integral, weighted by $S_\phi$, over field configurations that obey the boundary condition, $\phi\Big|_{\Sigma}=\varphi$.  In order to do this, it is convenient to rewrite the integration variable as a classical field satisfying the equations of motion with the boundary conditions plus quantum fluctuations that vanish on $\Sigma$:
\beq
\phi=\phi_c+\chi\qquad\qquad\Box\phi_c=0\qquad\qquad\phi_c|_{\Sigma}=\varphi(\vec x)\qquad\qquad \chi|_{\Sigma}=0.
\eeq
For completeness, this is reviewed in great detail in Appendix \ref{transamapp}.  The result is the correct ground state wave functional.  Additionally by considering the boundary of the manifold to consist of two surfaces, $\Sigma_\pm$, we show in the same appendix that the path integral weighted by $S_\phi$ produces the correct transition amplitudes.

\subsection{States}\label{sectarbstates}
As mentioned above, the procedure for constructing states from the path integral is simple. Let us first recall the construction of the ground state wave functional. Consider the \emph{Euclidean} path integral $Z[M_{-}; \varphi]$ on the lower-half space $\tau \leq 0$ (where $\tau$ is Euclidean time), with the boundary conditions $\phi(0,\vec{x}) = \varphi(\vec{x})$, and $\lim_{T \to \infty} \phi(-T,x) = 0$. From the standard time-slicing construction of this path integral,  we have
\beq
Z[M_-;\varphi ] = \lim_{T \to \infty} \left\langle \varphi(\vec{x}) \left| e^{-\int_{-T}^0d\tau \hat{H}} \right| 0\right\rangle
\eeq
where note that $|0\rangle $ is not the vacuum, but corresponds to the field configuration $\phi (\vec{x})=0$. However, by expanding $|0\rangle$ in terms of energy eigenstates, we see that the limit $T \to\infty$ isolates the vacuum state inside this expansion, and so we obtain\footnote{The projection onto the ground state pertains only when there exists a small gap, e.g., a mass term $\mu^2\phi^2$ in the action (and the state $|0\rangle$ has non-zero overlap with the vacuum).  Such a gap of course breaks the conformal invariance of the theory, so one might worry that this interferes with the renormalization group flow.  We will see later that $\mu^2$ is a special case of a bi-local source whose RG equations we will derive generically.  It will then be clear that the limit $\mu^2\rightarrow0$ is consistent with renormalization.}
\beq
Z[M_-;\varphi ] \sim  \left\langle \varphi(\vec{x}) |  \Omega \right\rangle
\eeq
where the overall normalization factor will be fixed shortly. So  we deduce that the Euclidean path integral over the lower half-space constructs the ground state wave functional $\left\langle \varphi(\vec{x}) |  \Omega \right\rangle$ (see fig 2(a)). 
\begin{figure}[!h]
\centering
\begin{tabular}{ l l l l}
\includegraphics[width=7cm]{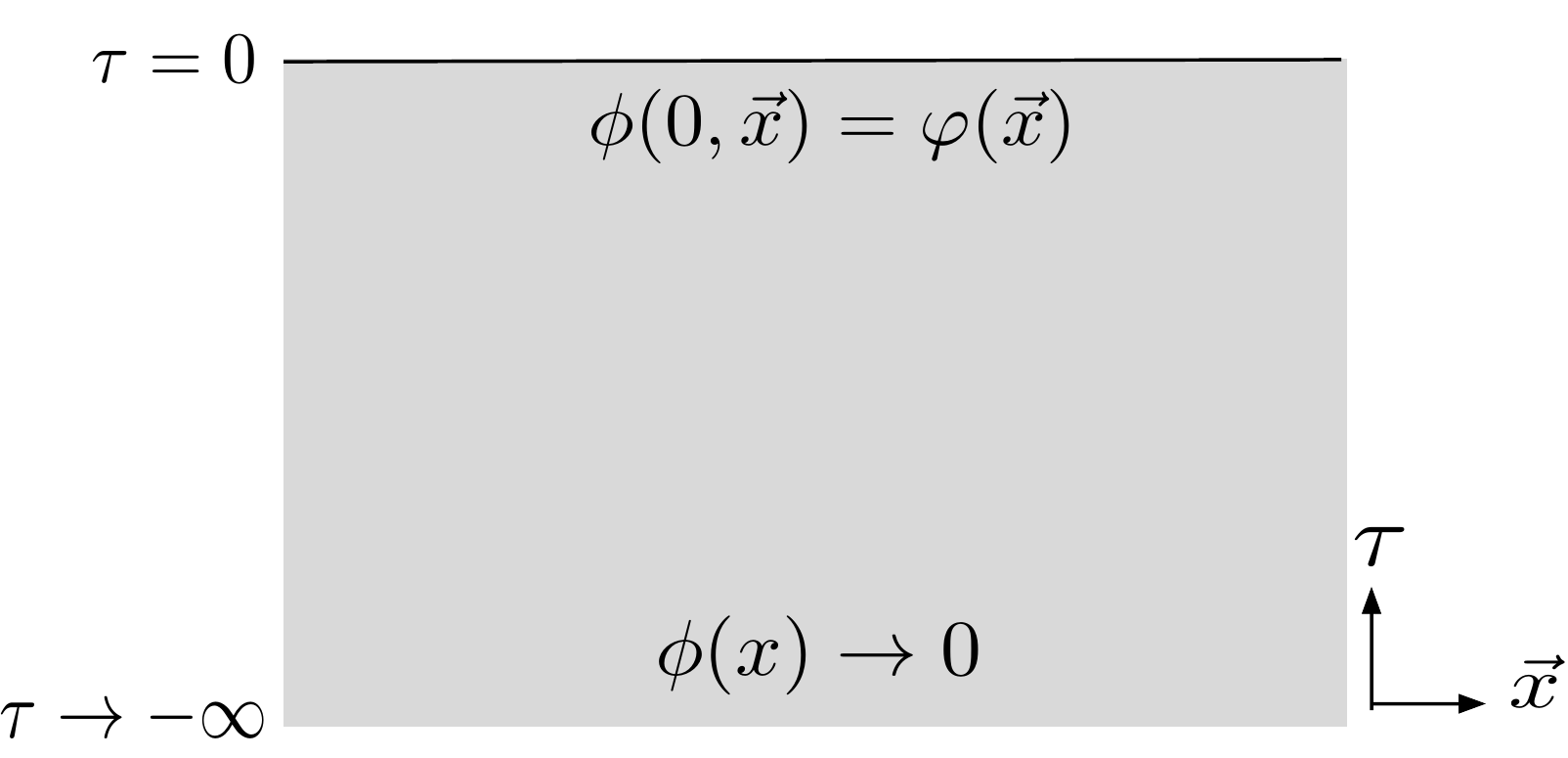} & 
\includegraphics[width=7cm]{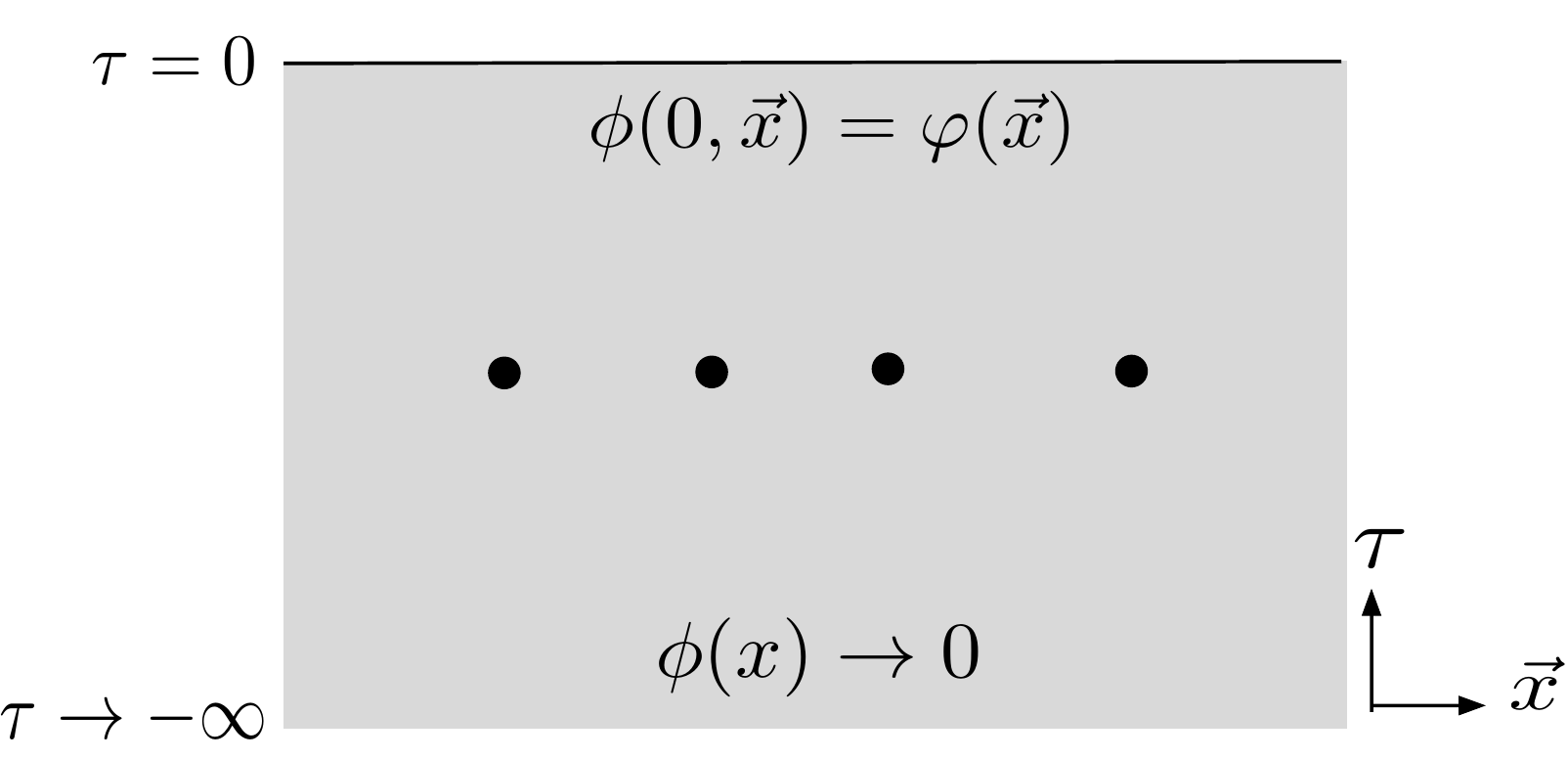}
\end{tabular}
\caption{\small{\textsf{(a) The ground state wave functional is given by the Euclidean path integral on the lower-half space. (b) Excited states can be constructed by operator insertions in the path integral.}} }
\label{fig:half}
\end{figure}

\begin{figure}[!h]
\centering
\begin{tabular}{ l l l l}
\includegraphics[height=5.7cm]{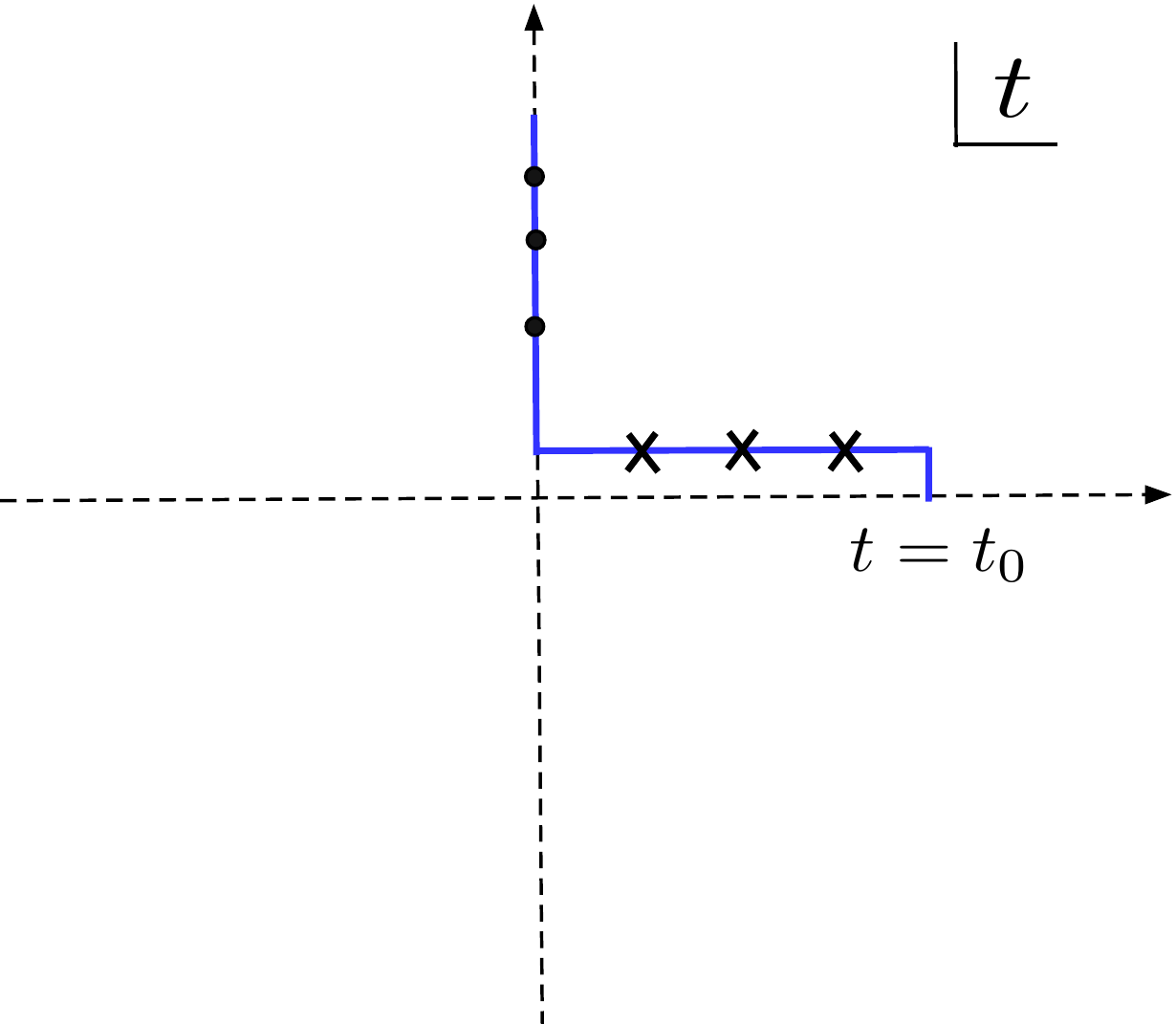} & \includegraphics[height=5.7cm]{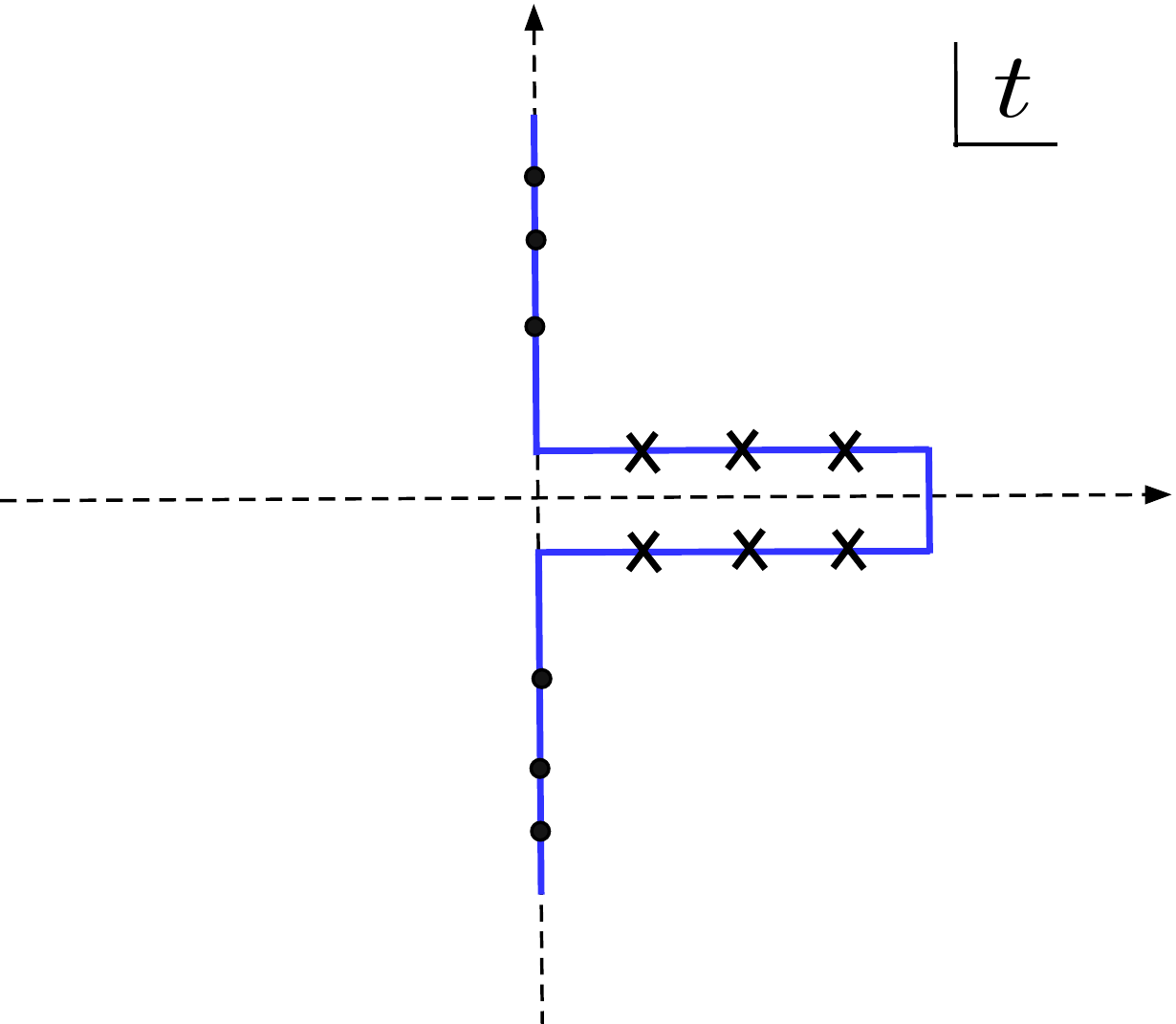}
\end{tabular}
\caption{\small{\textsf{(a) Operator insertions along the Euclidean branch of the contour prepares a state, while operator insertions in the real time calculate correlation functions in that state. (b) Operator and state norms are given by the path integral over the reflected contour.  Evolution by a small imaginary time $\delta$ ensures these norms do not suffer from contact divergences of operators acting at the same spacetime point.}} }\label{fig:RT.pdf}
\end{figure}

There are several directions in which we can generalize this construction. Firstly, we can construct excited states; in this paper, we will be interested in states of the form
\beq\label{states}
|\psi\rangle = e^{-\delta \hat H} \hcO_1(0,\vx_1)\hcO_2(0,\vec{x}_2)\cdots \hcO_n(0,\vx_n) |\Omega \rangle
\eeq
where the operators $\hcO_i$ are arbitrary, single trace $O(N)$-singlet operators (i.e., conserved currents of arbitrary spin and their descendants). In order to ensure the normalizability of these states, we will always take $\delta > 0$. These states can be constructed by performing the path integral on the lower half space, with the appropriate operator insertions at $\tau=-\delta$ (see Fig. \ref{fig:half}(b)). In fact, a convenient way to deal with these states in terms of path integrals is to package them in a \emph{generating functional}\footnote{As we will see shortly, potential divergences from the $\phi-\phi$ OPE are cancelled when we normalize these states.}
\beq
|\psi [B]\rangle =  \mathcal{T}_E e^{ -\frac{1}{2}\int_{M_-} d^dx\int_{M_-}d^dy\,\hat{\phi}^a(x)B(x,y)\hat{\phi}^a(y)} |\Omega\rangle
\eeq
where $\mathcal{T}_E$ denotes Euclidean time-ordering, and the bi-local source $B(x,y)$ can be thought of as taking the form of a ``differential operator''
\beq
B(x,y) = \sum_{s=0}^{\infty} B^{(s)}_{\mu_1\cdots \mu_s}(x) \pa^{\mu_1}\cdots \pa^{\mu_s} \delta^d(x-y).
\eeq
It should be clear from the time-slicing construction that $\mathcal{T}_E$ allows us to write the state $\psi[B]$ as the Euclidean path integral on the lower half-space, but with the action deformed by the source term
\beq
\langle \varphi(\vx)|\psi[B]\rangle = \int^{\phi(0,\vec{x})= \varphi(\vx)}_{\phi(-\infty,\vec{x})\to 0} [\mathscr{D}\phi]\;e^{-S},\;\;\;S = S_\phi +   \frac{1}{2}\int_{M_-} d^dx\int_{M_-}d^dy\,\phi^a(x)B(x,y)\phi^a(y)
\eeq
A second direction in which we can generalize is to consider the real-time evolution of the excited states discussed above, namely
\beq
|\psi(t_0)\rangle = e^{-\delta \hat H -it_0\hat H} |\psi \rangle
\eeq
This is easily accomplished by performing the path integral in complex time, along the contour $C$ shown in Fig. \ref{fig:RT.pdf} (with $t=-i\tau$ along the Euclidean section of the contour). In fact, it is a simple matter to generalize the generating functional of states to real time by extending the source $B$ along this contour -- we will refer to this generating functional as $|\psi_C[B]\rangle $. The utility of defining this generating functional is that it allows us to study the evolution of the more general class of states which are created by the contour-ordered action of operators, i.e.
\beq
|\psi(t_0)\rangle_C =  e^{-\delta\hat H-it_0\hat H}\mathcal{T}_C[ \hcO(t_1,\vec{x}_1)\cdots \hcO(t_n, \vec{x}_n)]|\Omega \rangle .
\eeq
The more general situation of out-of-time-order action of operators (in real time) requires a more complicated contour, and will not be considered in this paper.

Finally, let us discuss the norm of the state $|\psi_C[B]\rangle $. The dual $\langle \psi_C[B]|$ involves a reverse of contour ordering and so the combined contour over which the path integral is performed is the time-contour $\tilde{C}$ shown in figure \ref{fig:RT.pdf}(b). Therefore, the norm is given by the partition function of the theory for the time-contour $\tilde{C}$, but where the source $B$ is taken to satisfy the reflection symmetry
\beq
B(t_x,\vec{x}; t_y, \vec{y}) = B^*(t_x^*,\vec{x}; t_y^*,\vec{y})
\eeq
For the most part, we will find it convenient to divide out by the normalization and deal with the normalized generating functional
\beq
|\Psi_C[B]\rangle = \frac{1}{Z^{1/2}_{\tilde{C}}[B]}  |\psi_C[B]\rangle,\;\; Z_{\tilde{C}}[B] = \langle \psi_{C}[B] | \psi_C[B]\rangle
\eeq
It is worth emphasizing that the state $|\Psi_C[B]\rangle$ does not generate {\it normalized} excited states directly:
\beq
-i\frac{\delta}{\delta B(x,y)}|\Psi_C[B]\rangle = \Big(\hat{\cal O}(x,y) -\mathrm{Re}\,\langle \hat{\cal O}(x,y)\rangle_B\Big) |\Psi_C[B]\rangle,
\eeq
is not normalized to unity.

\subsection{Background Symmetry}\label{sectbacksymm}
Before moving on to renormalization, let us address a large background symmetry group present in our description of wave functionals. We have established that the wave functional is the path integral over field configurations with Dirichlet boundary conditions on a constant time slice $\Sigma$:
\beq\label{gf1}
\langle \varphi(\vx)|\psi_C[B]\rangle = \int [\mathscr{D}\phi]^{\phi(t_0,\vec x)=\varphi(\vec x)}_{\phi(-\infty,\vec x)\rightarrow 0}\;e^{iS_C[\phi]},
\eeq
where the subscript $C$ on the action indicates that we integrate along the time-contour $C$. In this path integral, $\phi$ is an integration variable and so we are always allowed a field redefinition of the form\footnote{In order to make contact with more familiar background symmetries, it is useful to take $\mathcal{L}$ to be quasi-local:
$$\mathcal{L} (x,y) = \delta^d(x-y)+\zeta^{\mu}(x)\pa_{\mu}^{(x)}\delta^d(x-y)+\zeta^{\mu\nu}(x)\pa_\mu^{(x)}\pa_\nu^{(x)}\delta^d(x-y) + \zeta^{\mu\nu\lambda}(x)\pa_{\mu}^{(x)}\pa_{\nu}^{(x)}\pa_{\lambda}^{(x)}\delta^d(x-y)+\cdots, $$
where we see that the first non-trivial term is a diffeomorphism, while the higher terms are higher spin transformations.}
\begin{equation}
\phi'(x)=\int d^{d}y\,\mathcal L(x,y)\phi(y) \equiv\left(\mathcal L\circp \phi\right)(x), \label{ol20}
\end{equation}
where we have introduced the ``$\circp$" product notation to denote an integration over interior points $y$. For now, let us ignore any subtleties associated with the boundary at $t=t_0$; we will return to these shortly.  Following \cite{Leigh:2014qca}, we require $\mathcal L$ to satisfy 
\beq
(\mathcal L^T\circp \mathcal L)(x,y)=\delta^{d}(x-y), \label{ol2}
\eeq
and further require the source to transform as
\begin{equation}\label{Btrans}
B\rightarrow \mathcal L^{-1}\circp B\circp \mathcal L
\end{equation}
Additionally, we must also introduce another source $W_{\mu}(x,y)$ to act as a background connection, which transforms as
\begin{equation}\label{Wmutrans}
W_\mu\rightarrow \mathcal L^{-1}\circp W_\mu\circp \mathcal L+\mathcal L^{-1}\circp \left\lbrack\pa_\mu,\mathcal L\right\rbrack_{\circp}.
\end{equation}
extending $\pa_\mu$ to 
\begin{equation}
D_\mu(x,y)=\pa_\mu^{(x)}\delta^{d}(x-y)+W_\mu(x,y).
\end{equation}
With these conditions, it is easy to check that the bulk action is invariant\footnote{More precisely, the transformation \eqref{ol20} induces a Ward identity that relates the partition function evaluated at one value of $B,W_\mu$ to another related to it by (\ref{Btrans},\ref{Wmutrans}).} under \eqref{ol20}.  We will refer to the group of these background symmetries of the path integral as $O(L^2)$, because these are linear transformations on square-integrable fields which satisfy the orthogonality condition \eqref{ol2}.  As was discussed in \cite{Leigh:2014qca}, it is always possible to take $W_\mu$ to be a flat connection $W_{\mu}^{(0)}$:
\beq
\pa_{\mu}W^{(0)}_{\nu}-\pa_{\nu}W^{(0)}_{\mu} + \left[W^{(0)}_{\mu}, W^{(0)}_{\nu}\right]_{\circp} = 0.
\eeq
This is because any corrections to $W^{(0)}$ under the renormalization group flow which are not flat can always be absorbed into the other source $B$. This is a special property of the bosonic theory. In this sense, the flat connection $W^{(0)}$ is associated with the free fixed point, while the source $B$ should be thought of as deforming away from the fixed point.  From here on we will assume that the background connection is flat, but for tidiness, refrain from including the $``(0)"$ superscript. 

We must now confront the fact that since we are dealing with wave functionals, our spacetime manifold has a boundary, namely the Cauchy surface $\Sigma$ at $t=t_0$. 
Indeed, we have to be careful in how we treat $\mathcal L$ as we approach $\Sigma$.  For example, in a derivative expansion the first non-trivial term acts as a diffeomorphism on $\phi$. We want to restrict to diffeomorphisms which preserve the Cauchy surface; similar remarks apply to higher spin transformations.\footnote{Additionally, as we discussed above, we have tuned our sources to zero within a neighborhood of width $\delta$ around $\Sigma$. We still wish to keep this buffer zone around $\Sigma$, for normalizability.}  This means we should restrict $O(L^2)$ transformations to become bi-local only in the spatial directions within the $\delta$ neighborhood of $\Sigma$:
  \begin{equation}
  \lim_{t \to t_0} \mathcal L(t,\vec x;t',\vec x')= \ell(t,\vec x;\vec x')\delta(t-t').
  \end{equation}
We will call this subgroup $O_{\{\Sigma\}}(L^2)$ to make explicit its dependence on the constant time boundary.  This subgroup then acts on the boundary values by
  \begin{equation}
  \varphi'(\vec x)=\int d^{D}\vec x'\ell(t_0,\vec x;\vec x')\varphi(\vec x') \equiv \left(\left.\ell\right|_{t_0}\pdot\varphi\right)(\vec x).
  \end{equation}
where the ``$\pdot$" product denotes an integration along the spatial coordinates of $\Sigma$.  In allowing $O_{\{\Sigma\}}(L^2)$ to act on the boundary values of $\phi$, we should be careful to covariantize the boundary term in the action (see \eqref{Fullaction}):
\begin{align}
\int_{\Sigma}d^{D}\vec x\,\varphi \left.\pa_t\phi\right|_{t}\rightarrow&\int_{\Sigma}d^{D}\vec x\,d^{D}\vec x'\,\varphi (\vec x)\left(\pa_t\,\delta^{D}(\vec x-\vec x')+w_t(t_0,\vec x;\vec x')\right)\phi(t_0,\vec x')\nonumber\\
&\equiv \int_{\Sigma }d^{D}\vx\,\varphi (t_0,\vx)  (D_t \pdot \phi)(t_0,\vx),
\end{align}
where in the above expression we have required $W^{(0)}_\mu$ to become temporally local within the $\delta$ neighborhood of $\Sigma$:
\begin{equation} \label{c1}
\lim_{t \to t_0} W^{(0)}_\mu(t,\vec x;t',\vec x')=w_\mu(t;\vec x,\vec x')\delta(t-t')
\end{equation}

In addition to the symmetries discussed above, there is also a scaling symmetry, which is going to be relevant for the renormalization group.  Let us introduce a Weyl parameter $z$ and rescale the metric, $\eta_{\mu\nu}\rightarrow z^{-2}\eta_{\mu\nu}$.  At the same time we rescale the bi-local source $B\rightarrow z^{d+2}B$ and the connection $W_\mu\rightarrow z^d\,W_\mu$; note from \eqref{c1} that the boundary connection rescales as $w_\mu\rightarrow z^{d-1}\,w_\mu$.  The field theory action now reads
\begin{equation}
S_\phi+S_{source}=\frac{1}{2z^{d-2}}\int_M \phi(x)\circ D^2\circ \phi(x)+\frac{1}{2z^{d-2}}\int_{\Sigma}\varphi(\vx) \cdot D_t\cdot\phi(\vx) +\frac{1}{2z^{d-2}}\int_M \,\phi(x) \circ B\circ\phi(x).
\end{equation}
With these redefinitions, it is easy to see that 
\beq
\phi \to \lambda^{\frac{d-2}{2}}\phi,\;\; z \to \lambda z
\eeq
is a symmetry of the action.
Together with $O_{\{\Sigma\}}(L^2)$ transformations, we will sometimes denote the full group (including the above scaling symmetry) as $CO_{\{\Sigma\}}(L^2)$.  We can now write down a Ward identity for $CO_{\{\Sigma\}}(L^2)$ which succinctly encodes the background symmetries discussed above:
\begin{align} \label{WI}
\left\langle \lambda^{\frac{d-2}{2}}\ell\cdot\varphi \Big|\Psi[z,M,B]\right\rangle
&=J_{(\lambda,\ell)}^{-N/2}\Big\langle \varphi\Big | \Psi[\lambda^{-1}z,\lambda^{-1}M,\mathcal L^{-1}\circ B\circ \mathcal L]\Big\rangle
\end{align}
where by using the notation $ |\Psi[z,M,B]\rangle$, we have chosen to explicitly display the dependence of the state on the cutoff $M$ (to be introduced shortly) and the Weyl factor $z$. Additionally, $\mathcal{L}$ is an $O_{\{\Sigma\}}(L^2)$ transformation which approaches $\ell$ close to $\Sigma$, and  $J^N_{(\lambda,\ell)}$ is the Jacobian from the switching of integration variables ($\mathscr D(\lambda^{\frac{d-2}{2}}\ell\cdot\varphi)=J_{(\lambda,\ell)}^N\mathscr D\varphi$) in the normalization.

\section{The Exact Renormalization Group for states}\label{regularizationsect}
Our discussion up to now has been formal; in practice, one must define a regularization scheme such that the path integrals in question exist and can be evaluated.  In this paper, following \cite{Polchinski:1983gv,Leigh:2014tza,Leigh:2014qca}, we will use a regularization scheme that eliminates the effects of the high momentum modes in the path integral by including a smooth cutoff function $K(s)$ in the action, with
\begin{equation}
K(0)=1\qquad\qquad\lim_{s\rightarrow\infty}K(s)=0.
\end{equation}
We augment our action to 
\begin{equation}\label{regaction}
S_\phi=\frac{1}{2z^{d-2}}\int_M \phi(x)\;\circ \;K^{-1}\left(-\frac{z^2}{M^2}\vec D^2\right)\circ D^2\circ \phi(x)+\frac{1}{2z^{d-2}}\int_{\Sigma}\varphi(\vx)\cdot K^{-1}\left(-\frac{z^2}{M^2}\vec D^2\right)\cdot D_t \cdot\left.\phi\right|_{\Sigma}(\vx)
\end{equation}
so that field configurations with large \emph{spatial} momentum contribute a rapidly varying phase to the path integral (or along Euclidean sections of the contour, provide exponentially suppressed contributions to the path integral). It is important to note that in doing so we are not truncating 
 the Hilbert space, but we are limiting the dynamics of high momentum modes. We note that the class of cutoff functions we use in this paper is different from the cutoff functions appearing in previous works \cite{Leigh:2014tza,Leigh:2014qca} --- in particular we choose to regulate with respect to the spatial Laplacian, $\vec D^2$, as opposed to the full d'Alembertian, $D^2$.  Although we do not expect different cutoff functions to change the essential physics, there are classes of cutoff functions that are better suited for a given calculation.  A discussion of why the above cutoff is natural for studying states is given in appendix \ref{sectlorreg}, but broadly the reason is as follows: the Hilbert space at the free fixed point can be thought of as a tensor product of harmonic oscillators corresponding to each momentum mode $\vec{k}$:
\beq
\mathcal{H} = \otimes_{\vec{k}\in \mathbb{R}^{d-1}} \mathcal{H}_{\vec{k}}
\eeq
The above choice of cutoff function preserves this structure (were we to include time-derivatives inside the cutoff function, this would no longer be the case), but tunes the parameters of the high-energy oscillators ($\vec{k} \gg M/z)$ so as to exponentially suppress their dynamics. 
Furthermore, by excising time derivatives from the cutoff we can more cleanly implement a variational principle consistent with fixing field configurations at fixed times.  

Let us begin by recalling how the exact renormalization program works for the Euclidean partition function (the details of which can be found in \cite{Leigh:2014qca}).  First we start with the partition function as a functional of the bi-local source, $B$, and the source for the identity, $\mathcal U$.  This partition function is defined in the presence of a regulator with a cutoff, $M$, and a Weyl scaling parameter $z$ (defined in the previous section): $Z_E[z, M, B, \mathcal U]$.  The renormalization of the partition function is then a two step process:
\begin{itemize}
\item The first step is to lower the cutoff $M\rightarrow\lambda M$ with $\lambda<1$.  In Polchinski's exact renormalization this is done directly by changing the cutoff function; in the Wilsonian method this is analogous to integrating out the high frequency modes.  This is then interpreted as the partition function of an effective theory at lower momenta, with new values of the sources:
\begin{equation}
Z_E[z, M, B,\mathcal U]=Z_E[z, \lambda M, \tilde B, \tilde{\mathcal U}].
\end{equation}
The expressions for $\tilde B$ and $\tilde{\mathcal U}$ can be derived in detail by the methods in \cite{Polchinski:1983gv}.  
\item The next step is to perform a $CO(L^2)$ transformation, $\mathcal L$, to bring the cutoff back to $M$ while scaling the Weyl factor to $\lambda^{-1}z$.  Since, as discussed above, this is a redefinition of path integration variables, this is also an identity:
\begin{equation}
Z_E[z,\lambda M,\tilde B,\tilde{\mathcal U}]=Z_E[\lambda^{-1}z, M, \mathcal L^{-1}\circ\tilde B\circ\mathcal L, \hat{\tilde{\mathcal U}}]\equiv Z_E[\lambda^{-1}z, M, B(\lambda^{-1}z), \mathcal U(\lambda^{-1}z)],
\end{equation}
(here $\hat{\tilde{\mathcal U}}$ allows for the possibility of a $CO(L^2)$ anomaly).  
\end{itemize}

Now we are interested in adapting the above discussion to the ERG flow of states or wave functionals, as opposed to partition functions. Since we're interested in the general class of excited states \eqref{states}, it is convenient to formulate this discussion in terms of the generating functional of states $\Psi_C[B]$, as has been explained previously. To be precise, what we are interested in here is the flow in the space of states of the undeformed CFT; that is, we are thinking of the states themselves, rather than correlation functions of operators in non-trivial states. This distinction is important conceptually, but in practice the difference between sources for operators and sources for operators that occur in the definition of a state amounts to where the sources occur along the contour. Locality in time along the contour is maintained throughout. These issues are discussed further in appendices.

 How should one formulate the renormalization group equation for the generating functional of states? Since $\Psi_C[B]$ can be written as a path integral along the appropriate time contour, we expect that the renormalization can be carried out by following the same steps as in the case of the Euclidean partition function; indeed, this is what we will do. The corresponding path integral will give us a one-parameter $\Psi_C[z,B]$, which we treat as the \emph{Wilsonian effective generating functional} of states. 
Thus, ERG for states can again be stated as a two-step process:

\textbf{Step 1: Lower the cutoff}
\\\\
To be explicit, recalling eq. \eqref{gf1}, the generating functional of states is given by the path integral
\begin{equation}
\Psi_C[z,M, B,\varphi]=\left\langle \varphi | \Psi_C \right\rangle = \mathcal N\int\left[\mathscr D\phi\right]^\varphi\exp\left(iS_\phi+iS_{source}\right)
\end{equation}
where $S_\phi$ is defined as in eq. (\ref{regaction}). 
From here we can compute $M\frac{\pa}{\pa M}\Psi$ from the standard Polchinski formalism, the only subtlety arising from treating the boundary terms carefully.  This is done in detail in appendix \ref{ERGdetapp} and we will simply quote the result here:
\begin{align}\label{lowercutoff} 
M&\frac{\pa}{\pa M}\Psi[z,M,B,\mathcal U;\varphi]\nonumber\\
&=\left(-z\Tr_{\Sigma\times C}\left(B\circ\Delta_B\circ B\circ\frac{\delta}{\delta B}\right)-z\frac{N}{4}\Tr_\Sigma\left(\Delta_\Sigma\right)-\frac{z}{2}\int_\Sigma\varphi \cdot \Delta_\Sigma\cdot\frac{\delta}{\delta\varphi}\right)\Psi[z,M, B, \mathcal U; \varphi].
\end{align}
In this formula we have defined 
\beq
\Delta_B=\left(D^2\right)^{-1}\circ \frac{M}{z}\frac{d}{dM}K,
\eeq
namely, the derivative of the field theory two-point function with respect to $M$, and
\beq
\Delta_\Sigma= \left. K^{-1}\frac{M}{z}\frac{d}{dM}K\right|_\Sigma
\eeq
is a boundary kernel.  Note that, since $K$ involves a flat connection, there is no ordering ambiguity in these definitions.\footnote{We note that a term of the form $\Tr_{\Sigma\times C}\left(\Delta_B\circ B\right)$ that was present in previous ERG calculations \cite{Leigh:2014tza,Leigh:2014qca}, is cancelled due to the normalization of the wave functional.  The normalization is additionally responsible for the appearance of the $\Tr_\Sigma(\Delta_\Sigma)$ term which arises from an integration by parts inside of the $\int[\mathscr D\varphi]$ integral.}

\textbf{Step 2: Scale transformation}\\\\
The second step is now to raise the cutoff by performing an infinitesimal $CO_{\{\Sigma\}}(L^2)$ transformation with $\lambda=1-\varepsilon$.  We parameterize this transformation by $\mathcal L=1+\varepsilon z\,W_z+O(\varepsilon^2)$ in the interior and $\ell =1+\varepsilon z\,w_z+O(\varepsilon^2)$ on $\Sigma$.
Using the Ward identity \eqref{WI}, we then obtain
\begin{empheq}[box=\widefbox]{align}\label{totalzexcstate}
z\frac{\pa}{\pa z}\Psi=\left(z\Tr_{\Sigma\times C}\left(\left([W_z,B]_\circ+B\circ \Delta_B\circ B\right)\circ\frac{\delta}{\delta B}\right)+z\frac{N}{2}\Tr_\Sigma g+z\int_\Sigma\varphi\cdot g^t \cdot\frac{\delta}{\delta\varphi}\right)\Psi.
\end{empheq}
where we have defined the bi-local kernel 
\begin{equation} \label{g}
g(z;\vec x,\vec y):=\left(\frac{1}{2}\Delta_\Sigma+w_z\right)(\vec x,\vec y)
\end{equation}
The trace over $\Sigma$ of $w_z$ arises from the Jacobian of the infinitesimal $CO_{\{\Sigma\}}(L^2)$ transformation. From the above equation we see that $\mathcal U$ plays only a spectator role in the ERG equation; from here on, we will drop it from the notation. We now want to understand the meaning of the various terms appearing in the above equation. We remark in passing that whereas in the case of the partition function, the RG principle is often stated as the partition function being independent of the value of the cutoff $M$, in the case of states, we simply organize the calculation in such a way as to eliminate derivatives with respect to $M$. Later we will see that this leads to an interpretation in which the ERG acts unitarily on states.

\subsection{Ground state}
Let us first investigate the above equation for the ground state wave functional.  This means we turn all sources off along the Euclidean section of our contour.  As we can see from \eqref{totalzexcstate}, this is consistent: setting $B=0$ completely eliminates the $\frac{\delta}{\delta B}$ terms in the differential equation and no extra operators are pulled down as we move along $z$:
\begin{equation}\label{pazGS}
z\frac{\pa}{\pa z}\Psi_\Omega[z,M,\varphi]=\left(z\frac{N}{2}\Tr_\Sigma\left(\frac{1}{2}\Delta_\Sigma+w_z\right)+z\int_\Sigma\varphi\cdot\left(\frac{1}{2} \Delta_\Sigma+w_z^{\text t}\right)\cdot\frac{\delta}{\delta\varphi}\right)\Psi_\Omega[z,M,\varphi].
\end{equation}
That is, the ERG flow induces no mixing of the ground state with other states at the UV fixed point.
Let us rearrange this equation slightly, to make it somewhat more transparent.  Recall that in the wave functional representation of a state, $\varphi$ and $\frac{\delta}{\delta\varphi}$ are the operators $\hat\phi$ and $\hat\pi$, respectively, acting on $\Psi_\Omega$:
\begin{equation}
\langle\varphi|\hat\phi(\vec x)|\Omega\rangle=\varphi(\vec x)\Psi_\Omega[\varphi]\qquad\qquad\qquad\langle\varphi|\hat\pi(\vec x)|\Omega\rangle=-i\frac{\delta}{\delta\varphi(\vec x)}\Psi_\Omega[\varphi].
\end{equation}
Together with \eqref{pazGS}, this then implies that the ground state satisfies the following flow equation in the $z$ direction
\beqn \label{quasidisent}
z\frac{d}{d z}\Big|\Omega(z)\Big\rangle &= & i\Big(\dis(z)+\dil(z)\Big) \Big|\Omega(z)\Big\rangle,\nonumber\\
\dis(z) &=&  \frac{z}{2}\left(\hat\pi\cdot \Delta_{\Sigma} (z)\cdot\hat\phi+\hat \phi\cdot \Delta_{\Sigma}^{\text t}(z)\cdot\hat \pi\right)\nonumber\\
\dil(z) &=&  \frac{z}{2}\left(\hat\pi\cdot w_z(z)\cdot\hat\phi+\hat \phi\cdot w_z^{\text t}(z)\cdot\hat \pi\right)
\eeqn
Both $\dis$ and $\dil$ are Hermitian operators. The operator $\dis(z)$ essentially damps out the dynamics of the UV modes, while preserving the direct product structure (in momentum space) of the vacuum. Of course, as we will see in the following section, generic excited states are not product states in momentum space -- in this case then, the operator $\dis(z)$ will play the role of a momentum space \emph{disentangler}, and will remove entanglement between modes above and below the scale $M/z$. Furthermore, by a judicious choice of the cutoff function (such as if we take it to be the exponential function) the operator $\dis$ may be thought of as a local operator in position space. On the other hand, the quasi-local operator $\dil$ implements the scale transformation which appears in the second step of ERG, in addition to a possible $O_{\{\Sigma\}}(L^2)$ transformation.

Interestingly, the differential equation \eqref{quasidisent} can be formally solved in terms of the path-ordered exponential of $\dis+\dil$:
\begin{equation}\label{GSpathordexp}
\Psi_\Omega[z_*,\varphi]=\Big\langle\varphi\Big|\Omega(z_*)\Big\rangle=\Big\langle\varphi\Big|\mathcal Pe^{\frac{i}{2}\int_\epsilon^{z_*}dz\int_\Sigma\left(\hat\pi\cdot g(z)\cdot\hat\phi+\hat\phi\cdot g^{\text t}(z)\cdot\hat\pi\right)}\Big|\Omega(\epsilon)\Big\rangle.
\end{equation}
where $|\Omega(\epsilon)\rangle$ is the UV ground state, and the kernel $g(z)$ was defined in \eqref{g}.  It is illuminating to write down this wave functional explicitly in the case where we choose to be a simple scale transformation at each $z$, $w_z(\vec x,\vec y)=-\frac{1}{z}\frac{(d-2)}{2}\delta^{D}(\vec x-\vec y)$.  Then solving \eqref{pazGS}, we obtain
\beq\Psi_\Omega[z,M,\varphi]={\det}_\Sigma\left(z^{2-d}K^{-1}|\vec D|\right)^{N/4}\exp\left(-\frac{1}{2z^{d-2}}\int\frac{d^{D}\vec k}{(2\pi)^{D}}\,\varphi(-\vec k)K^{-1}\left(\frac{z^2}{M^2}\omega_{\vec k}\right)\omega_{\vec k}\,\varphi(\vec k)\right)
\eeq
with $\omega_{\vec k}$ satisfying
\begin{equation}
\left(\left.\vec D^2\right|_\Sigma\cdot\psi_{\vec k}\right)(\vec x)=-\omega_{\vec k}^2\psi_{\vec k}(\vec x).
\end{equation}
for some complete set of eigenfunctions\footnote{e.g. in the gauge $w_\mu=0$, they can be chosen to be the plane waves $\psi_{\vec k}(\vec x)\sim e^{i\vec k\cdot\vec x}$ and $\omega_{\vec k}=\sqrt{\vec k^2}$.} $\psi_{\vec k}$ and $\varphi(\vec x)=\int\frac{d^{D}\vec k}{(2\pi)^{D}}\varphi(\vec k)\psi_{\vec k}(\vec x)$.  Similarly the determinant prefactor is defined by the product of $\omega_{\vec k}$.  Reassuringly, one can check that this wave functional defines the ground state at a scale $z$ by either explicitly doing the path integral defining it, or by canonical means, both of which are done in appendix \ref{transamapp}.

\subsection{Excited states}

Now let us investigate the flow of excited states, i.e. keeping $B\neq 0$, and make some comments about the nature of the RG equation, \eqref{totalzexcstate}.  We can rewrite this equation in terms of the newly defined Hermitian operators $\dis$ and $\dil$ as
\beq\label{statedepop}
z\pa_z \left|\Psi_C[B]\right\rangle =\left(-\Tr\,\boldsymbol{\beta}\circ\frac{\delta}{\delta B}+ i\dis + i\dil\right) \left| \Psi_C[B]\right\rangle
\eeq
where we have defined the bi-local beta function
\beq
\boldsymbol{\beta}(x,y) = -z\Big([W_z,B]_\circ +B\circ \Delta_B\circ B\Big)(x,y)
\eeq
However, we note that while the $\dis+\dil$ term in \eqref{statedepop} is a Hermitian operator defined without any explicit reference to the state on which it acts, the beta function term depends explicitly on the source $B$ preparing the state.  Since $|\Psi_C[B]\rangle$ is the generating functional for excited states, it should be clear from equation \eqref{statedepop} that the beta function term induces a mixing between the states $|\Omega\rangle,\, \mathcal{O}|\Omega\rangle,\; \mathcal{O} \mathcal{O} |\Omega\rangle, \ldots$ as we move in $z$.

A natural way to think of how RG is changing the initial state is to think of the generating functional of states $\Psi_C[B]$ as being a family of states parametrized by the source, $B$.   
In this language, we have a one-dimensional vector space (spanned by $\Psi_C$) fibered over the (infinite dimensional) manifold, $\mathcal M_B$, coordinatized by the bi-local sources $B(x,y)$. This base space is a generalization of the usual notion in RG of the space of couplings, here appearing as a parameterization of states. The beta function term in the RG equation for $\Psi_C$ can then be interpreted in terms of a flow along the vector field $\boldsymbol{\beta}$ on $\mathcal{M}_B$. In fact, if we interpret $z$ as parameterizing a curve\footnote{Note that we are using the notation ${\cal B}(z)$ to denote the analogue of a `running coupling', but we re-emphasize that here the sources are regarded as encoding non-trivial states of the CFT, rather than corresponding to turning on non-trivial couplings in the CFT.} 
\beq
\mathcal B: \mathbb R^+\rightarrow\mathcal M_B
\eeq
then $\Psi_C[z,\mathcal B(z)]$ flows along this particular curve\footnote{In this case the above equation admits another interesting interpretation -- as described previously, the state $\Psi_C[B]$ defines a line bundle over $\mathcal{M}_B$, equipped with a \emph{Berry connection} $\mathcal{A}$. In this context, the $\pa_z\mathcal B(z)$ term plays the role of the Berry connection pulled back to the RG curve through the space of sources.} via
\begin{equation}
\frac{d}{dz}|\Psi_C[z,\mathcal B(z)]\rangle=\frac{\pa}{\pa z}|\Psi_C[z,\mathcal B(z)]\rangle+\Tr\,\left(\pa_z \mathcal B(z)\circ\frac{\delta}{\delta \mathcal B(z)}\right)|\Psi_C[z,\mathcal B(z)]\rangle.
\end{equation}
We see from \eqref{statedepop} that along special curves solving 
\beq
z\pa_z\mathcal B=\boldsymbol{\beta}(\mathcal{B}), \label{brg}
\eeq
namely the integral curves of the vector field $\boldsymbol{\beta}$, the state $\Psi_C[z,\mathcal B(z)]$ has a particularly simple, \emph{unitary} flow equation:
\begin{equation}\label{excstpartrans}
z\frac{d}{dz}\left | \Psi_C[z,\mathcal B(z)]\right\rangle =
i\left(\dis + \dil\right) \left|\Psi_C[z, \mathcal{B}(z)] \right\rangle 
\end{equation}
of the same form as \eqref{quasidisent}.
\begin{figure}[H]
\centering
\includegraphics[width=.5\textwidth]{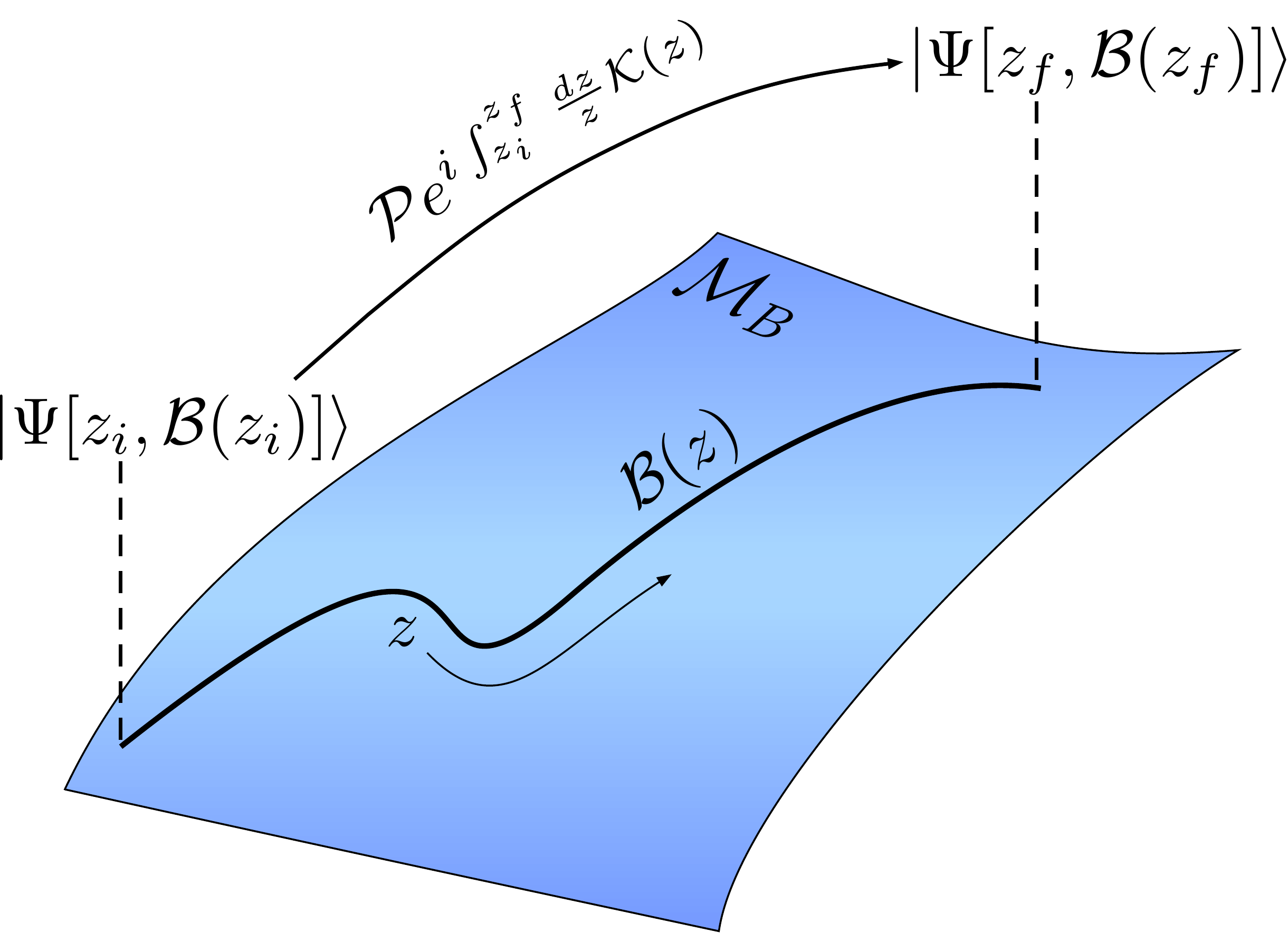}
\caption{Along integral curves $\mathcal B(z)$ of the beta function, $|\Psi[z;\mathcal B(z)]\rangle$ undergoes a unitary flow generated locally by the Hermitian operator $\mathcal{K} = \dis + \dil$.}
\label{sourcemanifoldfig}
\end{figure}
We note that in the picture we have presented, $\mathcal B$ prepares a state for the Hilbert space on $\Sigma$.  However, if we were to regard $\mathcal B(z)$ instead as a coupling for single-trace operators modifying the action of the theory and then $\Psi_C$ as the corresponding ground state, then \eqref{brg} is simply the ERG equation for the coupling $\mathcal{B}$.  In this case the flow of the ground state is captured entirely by the action of the Hermitian operators $\dis$ and $\dil$ {\em sans} any mixing with other states.  This suggests that taking the state to flow along integral curves of $\boldsymbol{\beta}$ is a natural requirement for ERG.  We argue that, in fact, unitary evolution of the state under ERG should be taken as an \emph{RG principle}.  In particular, the inner product of two unnormalized states is the Euclidean partition function.  Requiring that inner products remain unchanged through the flow in $z$ is then equivalent to requiring partition functions to be ERG invariants.   

We end this section by remarking that it is the structure of the ERG setup that allows us to require a unitary RG principle.  In particular the regulation of the theory is implemented not by excluding the high energy modes from the Hilbert space, but instead by altering the action (or equivalently the Hamiltonian) of the theory.  Because of this, lowering the cutoff does not change the size of the Hilbert space; the modes above the cutoff remain ancillary.  This is analogous to the ``exact holographic map" variant of entanglement renormalization \cite{Lee:2015vla}.  In the next section we will see that $\mathcal K(z)=\dis+\dil$ has a natural home in the language of unitary networks and we will expound upon the unitary structure of the ERG.

\section{The ERG as a tensor network}\label{cMERAsect}

We now want to better understand the right hand side of equation \eqref{statedepop}, and in particular the meaning of the hermitian operations appearing there.  To facilitate this, we will focus on the one parameter family of excited states, $\left|\Psi(z)\right\rangle \equiv \Big|\Psi[z,\mathcal B(z)]\Big\rangle$, evolving along the integral curve $z\pa_z\mathcal B=\boldsymbol{\beta}$ and which satisfies \eqref{excstpartrans}, or equivalently:
\begin{equation} \label{tn}
\left|\Psi(z)\right\rangle=\mathcal{P}\,e^{i \int_{z_{IR}}^{z}\frac{du}{u}\, \mathcal{K}(u)}\left|\Psi(z_{IR})\right\rangle,
\end{equation}
where $\mathcal{K} = \dis+\dil$. The scale-dependent unitary evolution under RG which we have found is not a novel concept.  This is in fact a central feature in (multi-scale) entanglement renormalization, MERA, which is a particular implementation of a \emph{tensor network} representation of certain special states (namely the ground state and low-energy excited states of a critical system). In MERA, a discrete system of finite size is ``evolved" step-by-step under the action of \emph{local} (where by local we mean action on nearest-neighbour sites) unitary\footnote{More precisely, these are taken to be isometries, but by adding ancillary degrees of freedom, it is possible to think of them as unitary operators.} operators, thus building a web of unitary operations culminating in an infrared state. The central precept of MERA is to choose the local unitaries carefully so as to spatially disentangle the state at each scale, thus leading to an IR state which is completely disentangled spatially. Continuous versions of MERA (cMERA) have also been proposed and studied \cite{Haegeman:2011uy,Nozaki:2012zj} for free quantum field theories, and are designed such that many of the above features carry over.

In our case, we see from equation \eqref{tn} that ERG naturally gives us a quasi-local continuum unitary evolution, reminiscent of cMERA. 
So what is the operator $\mathcal{K}$ doing at each step of ERG?  From its definition we see that it is the combination of a scale transformation $\dil$ and the operator $\dis$ which freezes the UV modes of the state above the given scale.  This is of course remarkably similar to the hermitian operation which is taken to generate the cMERA in free field theories -- the combined action of a scale transformation and a \emph{disentangler} which removes real-space entanglement at each RG step. So then the question which remains to be answered is whether we can interpret the operator $\dis$ which appears in ERG as a disentangler -- we claim that in fact $\dis$ plays the role of a \emph{disentangler in momentum space.} To illustrate this, we will consider a momentum-space reduced density matrix and track how it changes as momentum modes are traced over.

Before proceeding with this,  let us clarify the differences between the structure of the ERG setup and the familiar Wilsonian RG. In the latter, there is a hard cutoff in momentum space, and the RG is obtained by lowering the cutoff by explicitly integrating over the degrees of freedom within shells in momentum space. In Ref. \cite{Balasubramanian:2011wt}, it was shown that this leads to a reduced density matrix which is mixed. In the ERG setup on the other hand, all modes are present in the path integral but the high momentum structure of the Hamiltonian of the theory is modified by the presence of the cutoff function, and ERG corresponds to lowering the scale of the cutoff function. Nevertheless, we can still introduce the concept of a reduced density matrix in momentum space in the ERG setup, by explicitly tracing over modes with momenta above a given scale $\mu$. Such a process then can be expected to give rise to a path integral that is formally similar to that employed in Wilsonian RG. 

So we will now consider the momentum-space reduced density matrix of a specific excited state.  (Of course the ground state of the system is a product state in momentum space and so the calculation is trivial in that case; in order to understand the effect of $\dis$ we need to look at an excited state.)  For simplicity, let us pick the state created  by acting on the vacuum by a singlet operator which is local in position space\footnote{Since $\Big|\Psi_{\phi\phi}\Big\rangle$ involves a momentum integration, it is not a product state in momentum space.}:
\begin{equation}
\Big|\Psi_{\phi\phi}(z)\Big\rangle=\mathcal N\hat\phi(\vec x)\hat\phi(\vec x)\Big|\Omega(z)\Big\rangle.
\end{equation}
The matrix elements of the density matrix in the  $\varphi(\vec{p})$ basis are then:
\begin{equation}
\Big\langle\varphi_1(\vec p)\Big|\rho_{\phi\phi}\Big|\varphi_2(\vec q)\Big\rangle=\left|\mathcal N\right|^2\int_{\vec p_{1,2},\vec q_{1,2}}\!\!\!\!\!\!\!\!\!\!\!\!\!\!\!\!e^{i\sum(\vec p+\vec q)\cdot\vec x}\varphi_1(\vec p_1)\varphi_1(\vec p_2)\varphi_2(\vec q_1)\varphi_2(\vec q_2)e^{-\frac{1}{2z^{d-2}}\int_{\vec k}\left(\varphi_1(-\vec k)K^{-1}\omega_{\vec k}\varphi_1(\vec k)+\varphi_2(-\vec k)K^{-1}\omega_{\vec k}\varphi_2(\vec k)\right)}
\end{equation} 
Now we choose some reference momentum $\mu$, and calculate the reduced density matrix by tracing out all of the degrees of freedom with momenta $|\vec k|\geq\mu$, that is
\begin{equation}
\Big\langle\varphi_{1,<}(\vec p)\Big|\rho_{\phi\phi,<}\Big|\varphi_{2,<}(\vec q)\Big\rangle:=\int\prod_{|\vec k|\geq\mu}d\tilde\varphi_{>}(\vec k)\Big\langle\varphi_{1,<}(\vec p)\Big|\otimes\Big\langle\tilde\varphi_>(\vec k)\Big|\rho_{\phi\phi}\Big|\varphi_{2,<}(\vec q)\Big\rangle\otimes\Big|\tilde\varphi_>(\vec k)\Big\rangle.
\end{equation}
The Gaussian weight factorizes nicely under the splitting of the Hilbert space and the two-point and four-point integrations of $\tilde\varphi_>(\vec k)$ are performed easily.  Up to overall normalization we obtain
\begin{align}\label{reddenmat}
\rho_{\phi\phi,<}\sim& \left\{(N^2+2N)z^{2(d-2)}\int_{|\vec p|\geq\mu}\frac{K(\vec p)}{\omega_{\vec p}}\int_{|\vec q|\geq\mu}\frac{K(\vec q)}{\omega_{\vec q}}\right.\nonumber\\
&\left.\qquad+z^{d-2}\int_{|\vec q|\geq\mu}\frac{K(\vec q)}{\omega_{\vec q}}\int_{|\vec p_{1,2}|<\mu}\!\!\!\!\Big(N\varphi_{1,<}(\vec p_1)\varphi_{1,<}(\vec p_2)+N\varphi_{2,<}(\vec p_1)\varphi_{2,<}(\vec p_2)+4\varphi_{1,<}(\vec p_1)\varphi_{2,<}(\vec p_2)\Big)e^{i\sum\vec p\cdot\vec x}\right.\nonumber\\
&\qquad+\left.\int_{|\vec p_{1,2}|,|\vec q_{1,2}|<\mu}\!\!\!\!\!\!\!\!\!\!\!\!\!\!\!\!e^{i\sum(\vec p+\vec q)\cdot\vec x}\varphi_{1,<}(\vec p_1)\varphi_{1,<}(\vec p_2)\varphi_{2,<}(\vec q_1)\varphi_{2,<}(\vec q_2)\right\}e^{-\frac{1}{2z^{d-2}}\int_{|\vec k|<\mu}\left(\varphi_1K^{-1}\omega_{\vec k}\varphi_1+\varphi_2K^{-1}\omega_{\vec k}\varphi_2)\right)}
\end{align}
We see that the third line of \eqref{reddenmat} has the form of the original density matrix, restricted to the subspace of momenta less than $\mu$.  If the terms in the first and second lines were absent, then the reduced density matrix would be pure and identical in form to the original density matrix.  However, the presence of these terms indicates that the reduced density matrix is mixed and therefore this state would have non-zero entanglement entropy in momentum space.  Note though that each of these terms is weighted by factors of the cutoff function.
Thus, for scales $\mu^2>>\frac{M^2}{z^2}$, these integrals give vanishing contributions.  That is at scales greater than the cutoff, the reduced density matrix is essentially pure and the momentum space entanglement entropy at this scale vanishes.  Because the tracing out of $\tilde\varphi_>$ can be thought of in terms of  Wick contractions, this argument generalizes nicely to states formed by any polynomial of $\hat\phi\hat\phi$; higher order polynomials only give increasing powers of $K$.  
Because of the form of the cutoff function, the states with momenta much larger than the cutoff are disentangled from those with momenta less than the cutoff. In the ERG calculation, we note that the kernel $\Delta_\Sigma$ involved in the definition \eqref{quasidisent} of $\dis$ is determined by the cutoff function alone, and is peaked at the renormalization scale. Consequently, the operator $\dis$ implements the disentangling of the modes at momentum $\frac{M}{z}$ from low-energy modes. 

In terms of quantum information theory, the action of the unitary operator $\mathcal{P}\,e^{i \int_{z_{IR}}^{z}\frac{du}{u}\, \mathcal{K}(u)}$ is somewhat analogous to \emph{compression}; it packages the information in momentum space to modes below the effective cutoff scale $\frac{M}{z}$, with the density matrix above this scale being in a trivial product state. From this point of view, these UV modes are analogous to ancillary degrees of freedom required to implement unitary evolution in MERA. As we take $z$ larger, the subsystem carrying the entanglement gets increasingly smaller, but throughout the process the size of the Hilbert space remains the same and since the action is unitary, no information is lost.  

\begin{figure}[H]
\centering
\includegraphics[width=.5\textwidth]{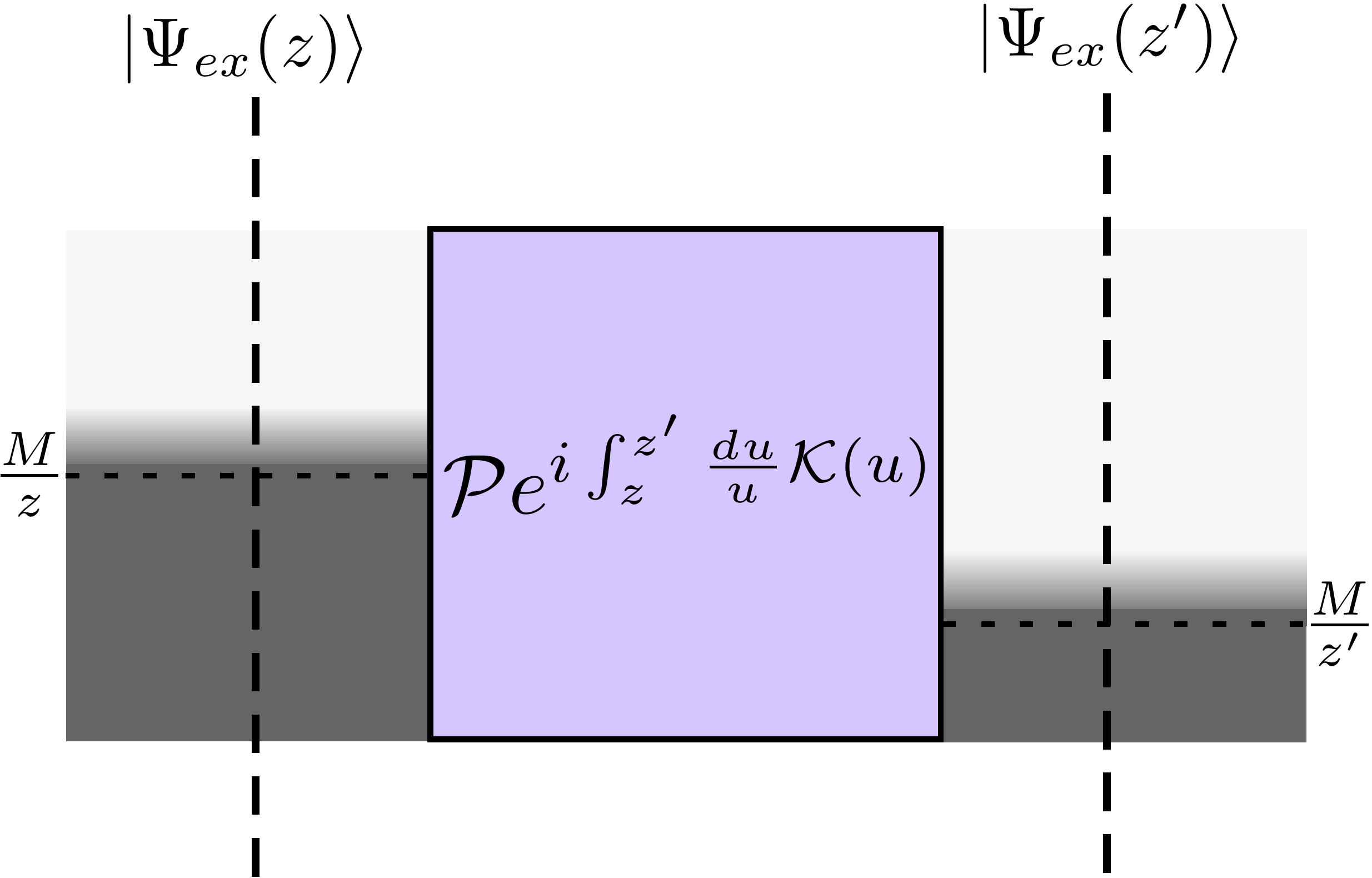}
\caption{The darker region indicates entanglement in momentum space.  The disentangler acts as a unitary channel that pushes this to lower and lower scales.}
\label{figcircuit}
\end{figure}

The arguments above only rely on the asymptotic behavior of $K$ at large momenta.  Of course, these asymptotics are fixed by requiring that loop integrals are UV finite via modifying the propagator:
\begin{equation}
\frac{i}{p^2+i\epsilon}\rightarrow \frac{iK\left(\frac{z^2\vec p^2}{M^2}\right)}{p^2+i\epsilon}.
\end{equation}  
That is, convergence of the field theory $n$-point functions naturally \emph{fixes} that the IR state has no momentum space entanglement. This is simply the organization given to us by ERG, given that there is a local (UV) free fixed point theory.  
We contrast this property of the ERG tensor network with the tensor network of cMERA.  In cMERA, a particular \emph{spatially} disentangled IR state is \emph{chosen}; the appropriate unitary transformations that connect it to the UV entangled state are determined by a variational principle.  
Now let us suppose that we could build the cMERA tensor network from the ERG process. As we saw above, the \disent\,is determined by a choice of cutoff function, so let us suppose that there existed a cutoff function that interpolates between the UV ground state at $z=\varepsilon<<1$ and the spatially unentangled IR ground state of \cite{Nozaki:2012zj}:
\begin{displaymath}
\xymatrix{
\Psi_\Omega[z,\varphi]=\mathcal N\exp\left(-\frac{1}{2z^{d-2}}\int_\Sigma\frac{d^{D}\vec k}{(2\pi)^{D}}\varphi(-\vec k)K^{-1}\left(\frac{z^2\omega_{\vec k}^2}{M^2}\right)\omega_{\vec k}\varphi(\vec k)\right)\ar[d]^{z\rightarrow\varepsilon<<1} \ar[dr]^{z\rightarrow\infty}&\\
\Psi_{UV}[\varphi]=\mathcal N_{UV}\exp\left(-\frac{1}{2\varepsilon^{d-2}}\int_\Sigma\frac{d^{D}\vec k}{(2\pi)^{D}}\varphi(-\vec k)\omega_{\vec k}\varphi(\vec k)\right)&\!\!\!\!\!\!\!\!\Psi_{IR}[\varphi]=\mathcal N_{IR}\exp\left(-\frac{1}{2z^{d-2}}\int_\Sigma\,\varphi(\vec x)\frac{M}{z}\varphi(\vec x)\right)}.
\end{displaymath}
It is easy to see that $\Psi_{IR}$ defines a state with no {\it spatial} entanglement: it is a product state in position space.  One possible cutoff function that would implement this could be taken to be
\begin{equation}
K_{cMERA}^{-1}\left(\frac{z^2\omega_{\vec k}^2}{M^2}\right)=\exp\left(\frac{1}{2}\log\left(\frac{M^2}{z^2\omega_{\vec k}^2}\right)\Theta\left(1-\frac{z^2\omega_{\vec k}^2}{M^2}\right)\right)=\left\{\begin{array}{cc}1&\qquad\frac{z^2\omega_{\vec k}^2}{M^2}<1\\
\frac{M}{z\omega_{\vec k}}&\qquad\frac{z^2\omega_{\vec k}^2}{M^2}\geq1\end{array}\right.
\end{equation}
Choosing such a cutoff function would then bridge the gap between the ERG and cMERA tensor networks.  More generally, requiring an ERG flow to the $\Psi_{IR}$ ansatz in the cMERA network requires the asymptotic behavior
 \begin{equation}
K_{cMERA}(s)=\left\{\begin{array}{cc}1&\qquad s<<1\\\sqrt s&\qquad s>> 1\end{array}\right.
\end{equation}
This does the opposite of what a regulator in a QFT is supposed to do --- it is enhancing the loop integrals in the large momentum regime.  The ``renormalization" of cMERA is then in this sense a qualitatively different procedure than renormalization in the Wilsonian or ERG sense.  

\section{Discussion}

In this paper we have applied the exact renormalization group procedure to path integral representations of states directly in continuum field theory.  In the process we have discovered that ERG naturally organizes itself into a continuous tensor network.  The nature of this tensor network is encoded in the Polchinski cutoff function and its purpose is to move entanglement in momentum space into an increasingly smaller domain of Hilbert space set by an effective renormalization scale, $M/z$.  Despite having a wide range of cutoff functions that one can use in ERG, the behavior of this network is set by the asymptotics of the cutoff function.  Ensuring that the correlators in the field theory are UV finite implies unambiguously that the IR state is one of vanishing momentum space entanglement.\\
\\
This should be contrasted with a very similar renormalization prescription for ground states of critical systems, cMERA.  In those prescriptions the IR state is chosen to be spatially unentangled and the tensors comprising the network are set by a variational principle.  We have argued that such renormalization prescriptions, while perfectly natural to consider in the broader context of unitary quantum circuits, are fundamentally different than renormalization in the Wilsonian sense.  They do not regulate UV divergences appearing in loop integrals. 
However, we note that in our discussion in the last section, it is the part of the cutoff function {\it above} the renormalization scale that differs between ERG and cMERA. Indeed, there is no doubt that the discrete version, MERA, where a fixed cutoff is provided by the lattice spacing, reproduces Wilsonian RG flow, which can be seen for example, in real space entanglement \cite{Evenbly:2008hca}.

In the recent literature, there has been a growing interest in the holographic interpretations of MERA \cite{Miyaji:2015fia, Lee:2015vla, Mollabashi:2013lya, Nozaki:2012zj} and it is clear that such schemes are capturing key features of holographic duality, particularly the nature of holographic entanglement entropy.  Early explanations of this involved interpreting the MERA network as living on and describing a discrete geometry of a constant-time slice of AdS \cite{Swingle:2009bg, Nozaki:2012zj}.  However, there is growing evidence that the connection between MERA and AdS geometry is not so direct.  For example,  the MERA tensor network fails consistency conditions for resolving length scales below the AdS radius \cite{Bao:2015uaa}.  Furthermore several groups have suggested a connection between MERA and de Sitter geometry on kinematic space
\cite{Czech:2015qta, Czech:2015kbp, Czech:2016xec}.  
In this paper, instead of extrapolating holographic geometry from a tensor network we have started with a system whose holographic geometry is well understood and extracted a tensor network from it.  Despite appearing superficially similar, this network has some important differences from cMERA.

There are several open questions at this stage.  The first is how do multi-trace interactions change the tensor network story, if at all?  It is well known that double trace deformations in the field theory amount to a change of boundary conditions in the bulk of AdS at large $N$, and the ERG framework can be extended to account for generic multi-trace interactions \cite{LP}.  It would be interesting to extend this program further to the context of the renormalization of non-trivial states.  

In this paper, we have laid the groundwork for the study of the renormalization of wave functionals in field theories. 
A natural extension would be to construct explicit reduced density matrices and hence to study the RG evolution of real space entanglement. 
Lastly, and perhaps most intriguingly, is the connection between ERG and quantum information.  
There have been many recent proposals for how quantum information encodes the key features of holographic duality from bulk reconstruction via quantum error correcting codes \cite{Almheiri:2014lwa, Pastawski:2015qua, Verlinde:2012cy}, threads of quantum information encoding the Ryu-Takayanagi formula for entanglement entropy \cite{Freedman:2016zud}, and the role of chaos and complexity in holographic geometry \cite{Stanford:2014jda, Roberts:2014ifa, Shenker:2013pqa, Shenker:2014cwa, Maldacena:2015waa}.  It is clear that there is a vast informational structure encoded in the holographic correspondence.  The tensor network constructed from ERG affords us only a glimpse at this structure.  Future work will explore and attempt to make this connection more explicit.

\vskip .5cm

{\bf Acknowledgements:}
Research supported in part by the U.S. Department of Energy contract DE-FG02-13ER42001.

\appendix\numberwithin{equation}{section}

\section{Appendix: Calculating the ground state wave functional}\label{transamapp}
Here we calculate the ground state wave functional in the cutoff theory at any $z$ by solving the free field path integral (i.e., with sources turned off).  Along the way we will derive an expression for generic transition amplitudes, with the ground state being given by a particular limit.  Afterwards we will verify this is the correct wave functional by writing down the Hamiltonian at a scale $z$ and showing it is the ground state, canonically.

Consider the transition amplitude between times $t_-$ and $t_+$ with fixed field configurations $\varphi_-$ and $\varphi_+$, respectively.  We will treat the time contour as being entirely along the real axis for simplicity, however this is not necessary: many of the details of the following will apply by taking a real parameterization of a complex time contour.  We will write this transition amplitude as a path integral with fixed field configurations
\begin{equation}
\Big\langle\varphi_+,t_+\Big|\varphi_-,t_-\Big\rangle\equiv \mathcal Z[\varphi_\pm]\equiv\mathcal N\int \left[\mathscr D\phi\right]^{\varphi_+}_{\varphi_-}e^{iS_\phi}
\end{equation}
with 
\begin{equation}
S_\phi=\frac{1}{2z^{d-2}}\int_{t_-}^{t_+}dt\int d^{D}\vec x\;\phi(t,\vec x)K^{-1}\left(-\frac{z^2}{M^2}\vec D^2\right)D^2\phi(t,\vec x)+\sum_\pm\pm\frac{1}{2z^{d-2}}\int_\Sigma d^{D}\vec x\;\varphi_{\pm}\cdot K^{-1}\left(-\frac{z^2}{M^2}\vec D^2\right)\cdot\left.D_t\phi\right|_{t_\pm}
\end{equation}
chosen to satisfy the variation principle with fixed field configurations.  The normalization of the path integral $\mathcal N$ is chosen so that two conditions are satisfied:
\begin{itemize}
\item Orthonormality:
\begin{equation}
\lim_{t_+\rightarrow t_-}\Big\langle\varphi_+,t_+\Big|\varphi_-,t_-\Big\rangle=\delta[\varphi_+-\varphi_-]
\end{equation}
\item Factorization by a complete set of states:
\begin{equation}\label{factorizationcond}
\int\mathscr D\tilde\varphi \Big\langle\varphi_+,t_+\Big|\tilde\varphi,\tilde t\Big\rangle\Big\langle\tilde\varphi,\tilde t\Big|\varphi_-,t_-\Big\rangle=\Big\langle\varphi_+,t_+\Big|\varphi_-,t_-\Big\rangle.
\end{equation}
\end{itemize}
Because this is a free field theory the path integral can be solved exactly by the field redefinition $\phi=\phi_c+\chi$ with $\phi_c$ the classical solution to the equations of motion subject to the boundary conditions $\varphi_\pm$ and $\chi$ the quantum fluctuations forced to zero at the boundary.  This shift leaves a path integral over $\chi$ with Dirichlet boundary conditions times a classical boundary action
\begin{align}
S_B=&\sum_\pm\pm\frac{1}{2z^{d-2}}\int_\Sigma d^{D}\vec x\;\varphi_\pm K^{-1}\left.D_t\phi_c\right|_{t_\pm}\nonumber\\
=&\frac{1}{2z^{d-2}}\int \frac{d^{D}\vec p}{(2\pi)^{D}}\frac{\omega_{\vec p}}{\sin(\omega_{\vec p}T)}K^{-1}\left(\frac{z^2}{M^2}\omega_{\vec p}^2\right)\left(\cos(\omega_{\vec p}T)\left(\varphi_+(\vec p)\varphi_+(-\vec p)+\varphi_-(\vec p)\varphi_-(-\vec p)\right)-2\varphi_+(\vec p)\varphi_-(-\vec p)\right).
\end{align}
where $T=t_+-t_-$.  The path integral over $\chi$ can evaluated using eigenfunctions of $D^2$.  Since $\chi$ is set to zero at $t_\pm$ the time-like momenta are valued over $k\in\mathbb Z_+$:
\begin{align}
\int[\mathscr D\chi]_{DBC}e^{iS_\chi}=&\prod_{\vec p}\prod_{k=1}^\infty\left(\frac{z^{2-d}K^{-1}}{2\pi}\frac{\pi^2k^2}{-T^2}\right)^{-N/2}\prod_{\vec p}\prod_{k=1}^\infty\left(1-\frac{\omega_{\vec p}^2T^2}{\pi^2k^2}\right)^{-N/2}\nonumber\\
=&\left(\int[\mathscr D\chi]_{DBC}\exp\left(\frac{i}{2z^{d-2}}\int_{t_-}^{t_+}dt\int d^{d-1}\vec x\;\chi K^{-1}(-\pa_t^2)\chi\right)\right)\prod_{\vec p}\left(\frac{\omega_{\vec p}T}{\sin(\omega_{\vec p} T)}\right)^{N/2}
\end{align}
where we have used the Euler Sine formula, $\sin(u)=u\,\displaystyle\prod_{k=1}^\infty\left(1-\frac{u^2}{\pi^2k^2}\right)$.  The transition amplitude is then
\begin{equation}
\mathcal Z[\varphi_\pm]=\mathcal N\left(\int [\mathscr D\chi]_{DBC}e^{-\frac{i}{2z^{d-2}}\int \chi K^{-1}\pa_t^2\chi}\right)\prod_{\vec p}\left(\frac{i2\pi T}{K^{-1}}\right)^{N/2}\times\prod_{\vec p}\left(\frac{z^{2-d}K^{-1}\omega_{\vec p}}{i2\pi\sin(\omega_{\vec p} T)}\right)^{N/2}\;e^{iS_B[\varphi_\pm]}.
\end{equation}
We've judiciously separated this into a product of two terms; the second is what correctly reproduces the delta function in the short time limit ($t_+\rightarrow t_-$ or $T\rightarrow 0$):
\begin{align}
\lim_{T\rightarrow0}\prod_{\vec p}\left(\frac{z^{2-d}K^{-1}\omega_{\vec p}}{i2\pi\sin(\omega_{\vec p}T)}\right)^{N/2}e^{iS_B[\varphi_\pm]}&=\lim_{T\rightarrow 0}\prod_{\vec p}\left(\frac{z^{2-d}K^{-1}}{i2\pi T}\right)^{N/2}\exp\left(\frac{i}{2z^{d-2}}\int\frac{d^{D}\vec p}{(2\pi)^{D}}K^{-1}\frac{(\varphi_+-\varphi_-)^2}{T}\right)\nonumber\\
&\equiv \delta[\varphi_+-\varphi_-].
\end{align}
This determines $\mathcal N$ to be
\begin{equation}
\mathcal N^{-1}=n_T^{-1}\left(\int [\mathscr D\chi]_{DBC}\exp\left(-\frac{i}{2z^{d-2}}\int_{t_-}^{t_+}dt\int d^{D}\vec x\,\chi K^{-1}\pa_t^2\chi\right)\right){\det}_\Sigma K^{N/2}
\end{equation}
with $n_T^{-1}={\det}_\Sigma(i2\pi T)$, a cutoff independent constant.  Given the explicit expression for the free transition amplitude
\begin{align}\label{explicitfreeamp}
\mathcal Z[\varphi_+,\varphi_-]=&\prod_{\vec p}\left(\frac{z^{2-d}K^{-1}\omega_{\vec p}}{i2\pi\sin(\omega_{\vec p}T)}\right)^{N/2}\nonumber\\
&\times\exp\left(\frac{i}{2z^{d-2}}\int \frac{d^{D}\vec p}{(2\pi)^{D}}\frac{\omega_{\vec p}}{\sin(\omega_{\vec p}T)}K^{-1}\left(\cos(\omega_{\vec p}T)\left(\varphi_+(\vec p)\varphi_+(-\vec p)+\varphi_-(\vec p)\varphi_-(-\vec p)\right)-2\varphi_+(\vec p)\varphi_-(-\vec p)\right)\right)
\end{align}
which is Gaussian in $\varphi_\pm$, it is easy to verify that the factorization condition, \eqref{factorizationcond}, is satisfied.  
\\\\
Given the discussion in section \ref{SingWFsect} we can use the above expression to determine the ground state wave functional by choosing a contour that begins and ends in positive and negative imaginary infinity, respectively.  For simplicity of computation, let's choose the contour  purely Euclidean, running from $t_i=iT$ to $t_f=-iT$ and evaluate this in the $T\rightarrow\infty$ limit.  Using \eqref{explicitfreeamp} we have
\begin{align}
\lim_{\beta\rightarrow\infty}\mathcal Z[\varphi_+,\varphi_-]=&\prod_{\vec p}\left(\frac{z^{2-d}K^{-1}\omega_{\vec p}}{\pi e^{2\omega_{\vec p}T}}\right)^{N/2}\!\!\!\!\!\!\!\!\times\exp\left(-\frac{1}{2z^{d-2}}\!\!\int\!\!\frac{d^{D}\vec p}{(2\pi)^{D}}K^{-1}\left(\frac{z^2}{M^2}\omega_{\vec p}^2\right)\omega_{\vec p}\left(\varphi_+(\vec p)\varphi_+(-\vec p)+\varphi_-(\vec p)\varphi_-(-\vec p)\right)\right)\nonumber\\
=&e^{-2T\,E_\Omega}\Psi^*_\Omega[\varphi_+]\Psi_\Omega[\varphi_-].
\end{align}
From here it is easy to isolate the expression for the ground state wave functional in the regulated theory:
\begin{equation}\label{appGSWF}
\Psi_\Omega[\varphi]=\prod_{\vec p}\left(z^{2-d}\pi^{-1}\omega_{\vec p}\,K^{-1}\right)^{N/4}\times\exp\left(-\frac{1}{2z^{d-2}}\int\frac{d^{D}\vec p}{(2\pi)^{D}}K^{-1}\left(\frac{z^2}{M^2}\omega_{\vec p}^2\right)\omega_{\vec p}\,\varphi(\vec p)\varphi(-\vec p)\right).
\end{equation}
Now let us write down the regulated Hamiltonian in the free theory:
\begin{equation}
\hat H(z)=\frac{1}{2}\int\frac{d^{D}\vec p}{(2\pi)^{D}}\left(z^{d-2}K\left(\frac{z^2}{M^2}\omega_{\vec p}^2\right)\hat \pi(-\vec p)\hat\pi(\vec p)+z^{2-d}K^{-1}\left(\frac{z^2}{M^2}\omega_{\vec p}^2\right)\omega_{\vec p}^2\,\hat\phi(-\vec p)\hat \phi(\vec p)\right)
\end{equation}
which is easily obtained from the Lagrangian and promoting the fields to operators.  The time dependent field operators can be expanded in the creation-annihilation basis
\begin{align}
\hat\phi(t,\vec x)=&\int\frac{d^{D}\vec p}{(2\pi)^{D}\,2\omega_{\vec p}}z^{\frac{d-2}{2}}K^{1/2}(\vec p)\left(\hat a_{\vec p}e^{-i\omega_{\vec p}t}+\hat a^\dagger_{-\vec p}e^{i\omega_{\vec p}t}\right)e^{i\vec p\cdot\vec x}\nonumber\\
\hat\pi(t,\vec x)=&-\frac{i}{2}\int\frac{d^{D}\vec p}{(2\pi)^{D}}\,z^{\frac{2-d}{2}}K^{-1/2}(\vec p)\left(\hat a_{\vec p}e^{-i\omega_{\vec p}t}-\hat a^\dagger_{-\vec p}e^{i\omega_{\vec p}t}\right)e^{i\vec p\cdot\vec x}
\end{align}
which diagonalize the Hamiltonian: $H=\frac{1}{2}\int_{\vec p}\left(\hat a^{\dagger}_{\vec p}\hat a_{\vec p}\right)+E_\Omega$.  Note that the factors of $z$ and $K$ in the field expansions ensure that $\hat a$ and $\hat a^\dagger$ have canonical commutation relations.  Now, using \eqref{appGSWF}, we verify that $|\Psi_\Omega\rangle$ is in fact annihilated by $\hat a$:
\begin{equation}
\Big\langle\varphi\Big|\hat a_{\vec p}\Big|\Psi_\Omega\Big\rangle=\left(z^{\frac{2-d}{2}}K^{-1/2}(\vec p)\,\omega_{\vec p}\,\varphi(\vec p)+z^{\frac{d-2}{2}}K^{1/2}(\vec p)\,\frac{\delta}{\delta\varphi(-\vec p)}\right)\Psi_\Omega[\varphi]=0.
\end{equation}
with energy 
\begin{equation}
\Big\langle\varphi\Big|\hat H(z)\Big|\Psi_\Omega\Big\rangle=\left(\text{vol}(\Sigma)\frac{1}{2}\int\frac{d^{D}\vec p}{(2\pi)^{D}}\;\omega_{\vec p}\right)\Psi_\Omega[\varphi]
\end{equation}
as expected.

\section{Foliations and Factorization}
Now we turn to a discussion of factorization of transition amplitudes.  In the canonical picture, the Hamiltonian flow naturally foliates the manifold into a collection of Cauchy surfaces, $\{\Sigma_t\}_{t\in[t_-,t_+]}$ defined by uniquely evolving the data at $t_-$ to $t$.  Along each $\Sigma_t$ we associate a Hilbert space spanned by $\left\{|\varphi,t\rangle, |\varpi,t\rangle\right\}$.  Of course, the usual maneuver of switching between Heisenberg and Schr\"odinger pictures is the statement that the Hilbert space at any given $t$ is isomorphic to any other $t_0$  via $|\varphi,t\rangle=e^{i\hat H(t-t_0)}|\varphi,t_0\rangle$; the label of $t_0$ can then be dropped. We can then ``cut open" transition amplitudes at any given time by an insertion of a complete set of states
\begin{equation}
\langle\varphi_+,t_+|\varphi_-,t_-\rangle=\langle\varphi_+|e^{-i\hat H(t_+-t_-)}|\varphi\rangle=\int[\mathscr D\tilde\varphi]\langle\varphi_+|e^{-i\hat H(t_+-\tilde t)}|\tilde\varphi\rangle\langle\tilde\varphi|e^{-i\hat H(\tilde t-t_-)}|\varphi_-\rangle.
\end{equation}
The statement of factorization is then that the transition amplitude over a time interval can be arbitrarily broken up into smaller transition amplitudes with the boundary conditions integrated over.  Given the description of transition amplitudes as path integrals with fixed boundary conditions then factorization tells us how to glue path integrals together: identify the boundary values of the fields and then integrate over the boundary fields.  Although this is fairly simple in the free field theory, let us describe how the bi-local sources glue across the boundary.\\
\\
Recall that we have regulated generic states and transition amplitudes by evolution with the free Hamiltonian by some $\delta$ away from $\Sigma_\pm$ where we define our boundary conditions.  With the bi-local sources tuned to zero within a width $2\delta$ neighborhood of the common boundary, $\Sigma_{\tilde t}$, there is no subtlety in doing the integration of the free theory at $\Sigma_{\tilde t}$.
After combining the actions, there is a strip of width $2\delta$ on which the bi-local source $B$ has no support.  This might seem at odds with the role $\delta$ played as an auxiliary regulator near the boundary but in fact, there is a natural method for washing this region out.\\\\
Each factored path integral carries an action of $O_{\{\Sigma_{\tilde t},\Sigma_\pm\}}(L^2)$ which is regarded as a redefinition of path integral variables.  These transformations were, generically, bi-local in time but were required to become temporally local as they approached either constant time boundary, $\Sigma_{\tilde t}$ or $\Sigma_\pm$. This group is natural in each factor; the constant time boundary breaks time translation leaving only spatial diffeomorphisms as symmetries of the action.  Additionally there were variational principles we were careful to preserve.  After integrating out $\tilde\varphi$ and gluing the path integrals along $\Sigma_{\tilde t}$ we now are free to enchance this symmetry $O_{\{\Sigma_+,\Sigma_{\tilde t}\}}(L^2) \times O_{\{\Sigma_{\tilde t},\Sigma_-\}}(L^2)\rightarrow O_{\{\Sigma_+,\Sigma_-\}}(L^2)$ to a group containing transformations that are bi-local even across $\Sigma_{\tilde t}$.  Recalling the action of $O_{\{\Sigma_+,\Sigma_-\}}(L^2)$ on a path integral with fixed boundary conditions
\begin{equation}
Z_\phi[z, M, B, \ell_+\cdot\varphi_+,\ell_-\cdot\varphi_-]=Z_\phi[z, M, \mathcal L^{-1}\circ B\circ \mathcal L,\varphi_+,\varphi_-]
\end{equation}
where $\ell_\pm$ is the value of $\mathcal L$ at $\Sigma_{\pm}$.  We see now that in this extended group we can we can regenerate bi-local sources in the gap by a simple change of path integral variables localized around $\Sigma_{\tilde t}$:
\begin{equation}
B'(t_1,\vec x;t_2,\vec y)=\int dt_3dt_4\int d^{D}\vec u\,d^{D}\vec v\,\mathcal L^{-1}(t_1,\vec x;t_3,\vec u)B(t_3,\vec u;t_4,\vec v)\mathcal L(t_4,\vec u;t_2,\vec y).
\end{equation}
$B'$ will have support in the $2\delta$ gap if $\mathcal L$ is chosen to have support in that region.  Thus, because of this enhanced background symmetry the full transition amplitude forgets about the $\delta$ regulator around $\Sigma_{\tilde t}$. For this reason, the choice of local boundary conditions for transition amplitudes does not actually interfere with factorization.
\begin{figure}[H]
\centering
\includegraphics[width=.9\textwidth]{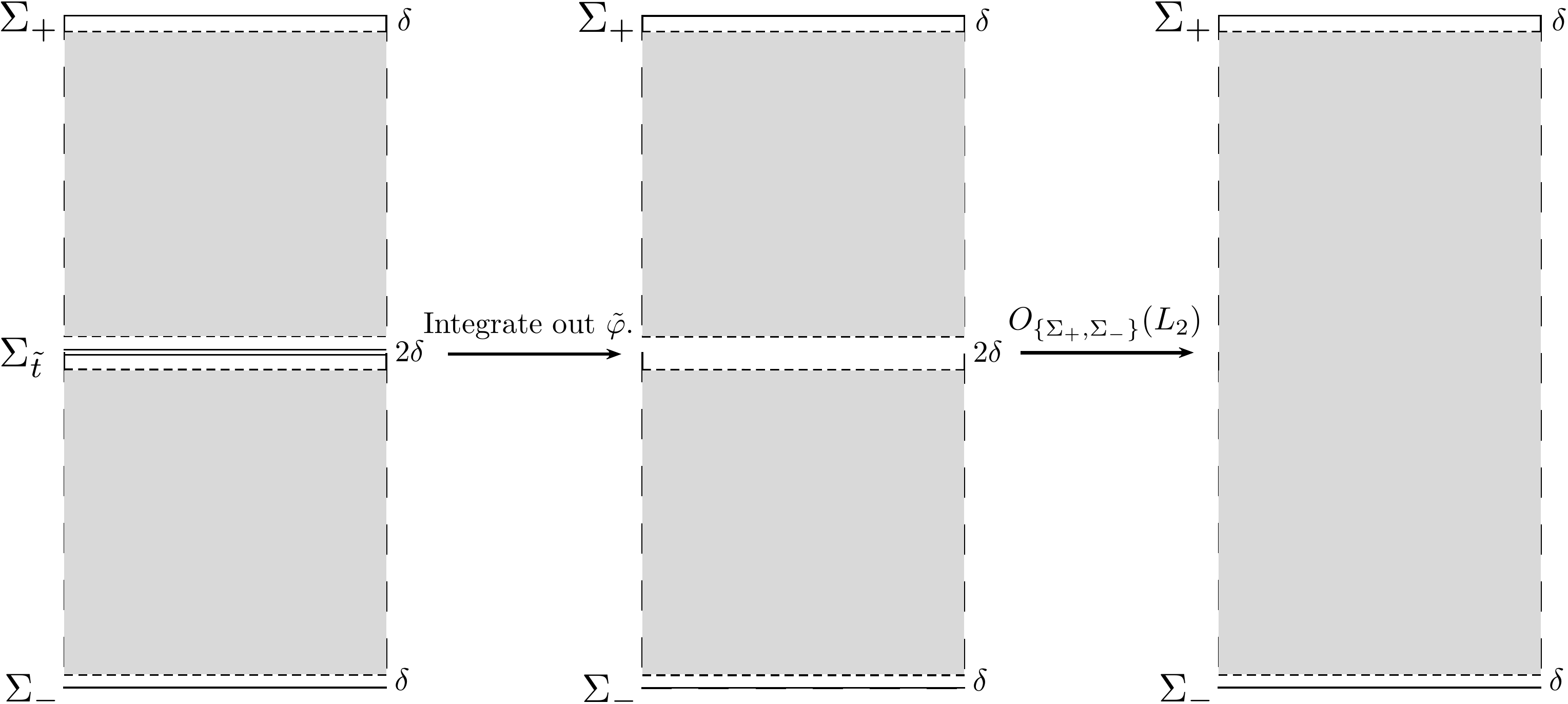}
\caption{Gluing path integrals with bi-local sources.  The shaded regions indicate the support of $B$.  After gluing there remains a region of width $2\delta$ along the contour on which $B$ has no support. However the action of a larger background symmetry generates new bi-local sources in this region.}
\label{figfactorization}
\end{figure}

\section{Regulators}\label{sectlorreg}
Let us review the class of regulators introduced in \cite{Leigh:2014qca, Leigh:2014tza}.  The path integrals in question were in fact Euclidean path integrals over the entire imaginary time interval, with the action possessing a global $SO(d)$ symmetry.  The cutoff functions then were chosen to preserve this symmetry; they were all of the form  
\begin{equation}
K_{Euc.}\left(-\frac{z^2}{M^2}\Box_{Euc.}\right)
\end{equation}
Let us recall briefly how this class of cutoff functions regulates the theory.  In the free theory, all $2N$-point functions can be evaluated via Wick's theorem and so we only need to show that the two-point function converges in the UV.  This is easy to see since we have augmented the propagator of the theory:
\begin{equation}
G(x,x')=\int\frac{d^dp_E}{(2\pi)^d}e^{ip_E\cdot (x-x')}K_{Euc.}\left(\frac{z^2}{M^2}p_E^2\right)\frac{1}{p_E^2}.
\end{equation}
So as long as  $\lim_{s\rightarrow\infty}K_{Euc}(s)\rightarrow0$ faster than $s^{1-d/2}$ this integral is UV convergent.
\subsection{Spatial Regulators and Complex Time Prescriptions}\label{sectspatreg}
Now let us consider going to the Lorentzian theory.  In this case using the same cutoff function involving $\Box$ will not do the trick;  the argument of the cutoff is no longer positive definite due to the signature:
\begin{equation}
K\left(-\frac{z^2}{M^2}\Box_{Lor.}\right)\rightarrow K\left(\frac{z^2}{M^2}\left(-\omega^2+\vec p^2\right)\right)
\end{equation}
Large momentum contributions can still contribute to loop integrals; if both $\omega$ and $\vec k$ are taken to be large then the argument of $K$ can still be small.  
A fix for this is to only regulate the spatial derivatives, taking a cutoff function
$
K\left(-\frac{z^2}{M^2}\vec\nabla^2\right).
$
This class of regulators has an additional appeal for calculating transition amplitudes and wave functionals defined on a Cauchy surface.  Firstly, in these situations, the global $SO(1,d-1)$ symmetry has been broken by constant time boundaries; there should be no particular reason to preserve this symmetry then in the action.  Secondly, and more importantly, there is a variational principle that we want to keep consistent.  That is if we fix field configurations on constant time boundaries then for any terms we include in the interior action, we must arrange a boundary action such that their sum variation is set to zero by the equations of motion and the boundary conditions.  By including a general functional of $\pa_t^2$ (which, in general, admits a power series of all orders) in the interior action there is no boundary action we can conjure to satisfy the variational principle. So in fact, the variational problem cannot be made well-posed in such a case; again, a regulator 
$K\left(-\frac{z^2}{M^2}\vec\nabla^2\right)$ sidesteps this problem.

Of course, whether or not a regulator makes sense variationally is moot if it doesn't properly regulate the theory.  Fortunately, the theory is properly regulated by choosing the appropriate time contour.  To see that this is so, note that a choice of time contour implies a contour, $C_\omega$, in the complex $\omega$-plane by requiring that the eigenfunctions of $\Box$ are complete along a real parametrization of $t$:
\begin{equation}
\int_{C_\omega} \frac{d\omega}{2\pi}\int\frac{d^{D}\vec k}{(2\pi)^{D}}e^{i\omega t-i\vec k\cdot\vec x}=\delta(t)\delta^{D}(\vec x).
\end{equation}
For instance, choosing $t=\tilde t(1-i\epsilon)$ with $\tilde t\in\mathbb R$ then $\omega$ must be $\tilde \omega(1+i\epsilon)$ with $\tilde \omega\in\mathbb R$ to counter act this:
\begin{equation}
\int_{C_\omega}\frac{d\omega}{2\pi}e^{i\omega t}=(1+i\varepsilon)\int_{-\infty}^\infty\frac{d\tilde\omega}{2\pi}e^{i\tilde\omega\tilde t}+O(\epsilon^2)=(1+i\varepsilon)\delta(\tilde t)+O(\epsilon^2)=\delta(t)+O(\epsilon^2).
\end{equation}
This is then the Feynman contour giving the time-ordered Green function.  Regulating the $\omega$ integral is unnecessary in the Lorentzian theory: its function is to enclose the poles at $\pm\omega_{\vec k}=\pm\sqrt{\vec k^2}$.  The propagator 
\begin{align}
G(t,\vec x; 0,\vec 0)=&\int_{C_\omega}\frac{d\omega}{2\pi}\int\frac{d^{D}\vec k}{(2\pi)^{D}}e^{i\omega t-i\vec k\cdot\vec x}K\left(\frac{z^2}{M^2}\vec k^2\right)\frac{1}{-\omega^2+\vec k^2}\nonumber\\
=&\frac{i}{2}\left(\theta(t)-\theta(-t)\right)\int\frac{d^{D}\vec k}{(2\pi)^{D}}K\left(\frac{z^2}{M^2}\vec k^2\right)\frac{1}{\sqrt{k^2}}e^{-i\vec k\cdot\vec x}
\end{align}
is UV convergent as long as $\lim_{s\rightarrow\infty}K(s)\rightarrow 0$ faster than $s^{1-d/2}$.  In the free theory this determines all of the higher point functions to be UV convergent as well.

It is natural of course to ask whether a non-Lorentz-invariant regulator creates a problem for the underlying Lorentz invariance of the theory. Of course, what we mean by Lorentz invariance here is the existence of a Ward identity satisfied by the partition function. In the case of a Lorentz-invariant regulator, this Ward identity is just a special case of the more general $O(L^2(\mathbb{R}^{1,d-1}))$ Ward identities. In the present case, it is much more involved; the specification of a non-Lorentz invariant regulator involves the choice of a space-like hypersurface, and so the Lorentz Ward identity relates partition functions (or transition amplitudes more generally) defined with distinct regulators. 


\section{Appendix: Details of ERG with fixed boundary conditions}\label{ERGdetapp}
We start with an important preliminary on the notation we will be using.  Consider a general functional $\mathcal F[\phi]$. If we have a variational principle that is consistent with fixing $\phi$ on $\Sigma$ then the general form of the variation of $\mathcal F[\phi]$ is 
\begin{equation}\label{generalvariation}
\delta \mathcal F=\int_C dt\int d^{D}\vec x\,\, \delta\phi(t,\vec x)\frac{\delta\mathcal F}{\delta\phi(t,\vec x)}_{bulk}+\int_{\Sigma}d^D\vec x\,\delta\phi(\vec x)\,\frac{\delta\mathcal F}{\delta\phi(x)}_{\Sigma}.
\end{equation}
The second term can come from the variation of a potential boundary action that $\mathcal F$ might contain \emph{plus} the integrations by parts needed to isolate $\delta\phi$ in the bulk integral if $\mathcal F$ contains derivative terms.  If after this process, the boundary integral is not of the above form but also contains derivatives of $\delta\phi$ then we say that $\mathcal F$ is not variationally consistent with fixing $\phi$ on $\Sigma$ in which case either extra boundary terms must be added to cancel these contributions or we must impose boundary conditions on $\mathcal F$ itself to ensure the coefficients are zero on the boundary.  From here on in this section we will assume that this tailoring has already been done.  We then take \eqref{generalvariation} as the definitions of $\frac{\delta\mathcal F}{\delta\phi}_{bulk}$ and $\frac{\delta\mathcal F}{\delta\phi}_{\Sigma}$.
\subsection{A Ward identity}\label{sectWardId}
Let us prove a simple Ward identity that we make use of in the renormalization procedure.  Suppose we want to consider the contour ordered correlation of a general operator of $\phi$, $\hat{\mathcal O}$ with $\varphi$ fixed on a (or possibly a disconnected set of) space-like boundary $\Sigma$.  In the case that the contour is Euclidean with $\Sigma$ consisting of a surface as $T\rightarrow\infty$ and a surface at $t=0$ this could be using $\hat{\mathcal O}$ to prepare a wavefunction, or if the contour takes an excursion along the real time axis this could be computing a real transition amplitude.  In this Ward identity we will be fairly agnostic about the specific contour $C$ and the specifics of $\Sigma$ and $\varphi$.\\\\
Now let us imagine a local shifting the field, $\phi(t,\vec x)\rightarrow \phi'(t,\vec x)=\phi(t,\vec x)+a f_{t_0,\vec x_0}(t,\vec x)$, where $a$ is an infinitesimal real number and $f_{t_0,\vec x_0}$ is a function with compact support about $(t_0,\vec x_0)$.  Inside of the path integral $\phi$ is a dummy variable and so the numerical result is unchanged by this field redefinition.  By choosing $(t_0,\vec x_0)$ sufficiently far away (or equivalently the support of $f_{t_0,\vec x_0}$ small enough) from $\Sigma$ this leaves the boundary conditions of the path integral unchanged:
\begin{align}
\int&[\mathscr D\phi]^{\varphi}\mathcal O[\phi]\,e^{iS[\phi]}e^{i\int_C B_i\mathcal O_i[\phi]}=\int\left[\mathscr D\phi'\right]^\varphi\,\mathcal O[\phi']e^{iS[\phi']}e^{i\int_C B_i\mathcal O[\phi']}\nonumber\\
&-a\int\!\!\left[\mathscr D\phi'\right]^\varphi\int f_{t_0,\vec x_0}\left\{\frac{\delta\mathcal O[\phi']}{\delta\phi'}_{bulk}+\mathcal O[\phi']\left(i\frac{\delta S[\phi']}{\delta\phi'}_{bulk}+iB_i\frac{\delta\mathcal O_i[\phi']}{\delta\phi'}_{bulk}\right)\right\}e^{iS[\phi']}e^{i\int_C \!\!B_i\mathcal O_i[\phi']}+O(a^2)
\end{align}
This holds for any function $f_{t_0,\vec x_0}$ localized away from $\Sigma$.  Taking this to have delta function support leaves us with a Ward Identity
\begin{equation}
\boxed{
\hat{\mathcal O}\,\hat{\frac{\delta S}{\delta\phi}}_{bulk}\;\sim\;-i\hat{\frac{\delta\mathcal O}{\delta\phi}}_{bulk}-\hat{\mathcal O}\,B_i\hat{\frac{\delta \mathcal O_i}{\delta\phi}}_{bulk}
}
\end{equation}
where ``$\sim$" denotes that this is an equality holding in contour ordered correlations functions and transition amplitudes.  The variation of the action with respect to the bulk field is, by definition, the equations of motion so this simply the familiar statement that in the quantum theory,  operators that vanish on-shell are redundant.  Though they may not be zero due to contact terms, they can always be written in terms of other operators.
\\\\
In particular, for the free $O(N)$ real scalar with higher spin operators sourced, this implies
\begin{equation}
\hat\phi(x)\pa_\mu\pa^\mu\pa_{\mu_3}\ldots\pa_{\mu_s}\hat\phi(y)\,\sim\,-iN\pa_{\mu_3}^{(y)}\ldots\pa_{\mu_s}^{(y)}\delta^d(x-y)-\sum_{s=0}^{\infty}\sum_{\{\nu\}}B^{\nu_1\ldots\nu_{s'}}\phi(x)\pa_{\mu_3}\ldots\pa_{\mu_s}\pa_{\nu_1}\ldots\pa_{\nu_{s'}}\phi(y)
\end{equation}
so that the traced higher spin currents are redundant: up to contact terms they can be expressed in terms of other higher spin currents.  This holds true along whichever time contour $C$  defines the path integral and regardless of boundary conditions defined on $\Sigma$.  Because of this, correlation functions containing two or more time derivatives can always be exchanged for spatial derivatives, guaranteeing that we can source \emph{all} operators in the single-trace spectrum without spoiling the variational principle of the interacting Lagrangian.

\subsection{ERG}
Let us now detail the exact renormalization of wave functionals.  The central object in the computation is the path integral defined along a time contour, $C$, and with the field fixed at $\varphi$ on $\Sigma$.  As we did in the previous appendix, we will treat $\Sigma$ as if it consisted of one space-like boundary at a fixed $\tilde t$ but the following results are easy to generalize to when $\Sigma$ has two components; one keeps track of the signs by the orientation of each component of $\Sigma$.  Let us denote this path integral by
\begin{equation}
\mathcal Z[C, \Sigma, \varphi, z, M, B]\equiv Z_\chi^{-1}\,Z_\phi
\end{equation}
where
\begin{equation}
Z_\chi\equiv \int[\mathscr D\chi]_{DBC}\exp\left(-\frac{i}{2z^{d-2}}\int_C \chi\circ K^{-1}\circ D_t^2\circ \chi\right),\qquad Z_\phi\equiv \int[\mathscr D\phi][ \mathscr D\rho]\exp\left(iS_\phi+iS_{source}+iS_\rho\right)
\end{equation}
are the path integrals we introduced in appendix \ref{transamapp}.  In particular $Z_\chi$ is there to ensure proper normalization.  The path integral over $\rho$ is arranged to enforce the boundary conditions of $\phi$.  The actions in $Z_\phi$ are
\begin{align}
S_\phi&=\frac{1}{2z^{d-2}}\int_C\phi\circ K^{-1}\circ D^2\circ\phi+\frac{1}{2z^{d-2}}\int_\Sigma\left.\left(\phi\cdot K^{-1}\cdot D_t\cdot\phi\right)\right|_{\tilde t}\nonumber\\
S_{source}&=\frac{1}{2z^{d-2}}\int_C\phi\circ B\circ \phi+i\,\mathcal U\nonumber\\
S_\rho&=\frac{1}{z^{d-2}}\int_\Sigma\rho\cdot\left(\phi(\tilde t)-\varphi\right)
\end{align}
Although we denote $S_{source}$ as an integration along the entire contour, $C$, we recall that we turn off the source $B$ within a time $\delta$ near the boundary, $\Sigma$.  Additionally we have included a source for the identity operator.  Let us pause quickly to note that given our definitions of $\frac{\delta}{\delta \phi}_{bulk}$ and $\frac{\delta}{\delta\phi}_\Sigma$ from the above section each of these actions have variations
\begin{align}
\frac{\delta S_\phi}{\delta\phi}_{bulk}&=\frac{1}{z^{d-2}}K^{-1}\circ D^2\circ\phi &\frac{\delta S_\phi}{\delta\phi}_\Sigma&=\frac{1}{z^{d-2}}\left.K^{-1}\cdot D_t\cdot \phi\right|_{\tilde t}\nonumber\\
\frac{\delta S_{source}}{\delta\phi}_{bulk}&=\frac{1}{z^{d-2}}\,B\circ \phi &\frac{\delta S_{source}}{\delta\phi}_\Sigma&=0\nonumber\\
\frac{\delta S_\rho}{\delta\phi}_{bulk}&=0 &\frac{\delta S_\rho}{\delta\phi}_\Sigma&=\frac{1}{z^{d-2}}\rho.
\end{align}
Instead of integrating by parts in the path integral, we will implement the ERG trick by repeated use of the Ward identity derived in \ref{sectWardId} which implies for the path integrals $Z_\chi$ and $Z_\phi$
\begin{align}\label{specificWardId}
\mathcal O[\chi]\circ K^{-1}\circ D_t^2\circ \chi\sim&\,iz^{d-2}\frac{\delta\mathcal O}{\delta\chi}_{bulk}\nonumber\\
\mathcal O[\phi]\circ K^{-1}\circ D^2\circ \phi\sim&\,-iz^{d-2}\frac{\delta\mathcal O}{\delta\phi}_{bulk}-\mathcal O[\phi]\circ B\circ \phi.
\end{align}
respectively.  Now we lower the cutoff in each path integral.  First let us do this for $Z_\chi$:
\begin{align}
M\frac{\pa}{\pa M}Z_\chi=&\int[\mathscr D\chi]_{DBC}\left(-\frac{i}{2z^{d-2}}\int\chi\circ M\frac{d}{dM}K^{-1}\circ D_t^2\circ \chi\right)e^{iS_\chi}\nonumber\\
=&\int[\mathscr D\chi]_{DBC}\left(\frac{i}{2z^{d-3}}\int\chi\circ K^{-1}\circ D^2\circ \Delta_B\circ K^{-1}\circ D_t^2\circ \chi\right)e^{iS_\chi}
\end{align}
We've defined the kernel
\begin{equation}
\Delta_B\equiv \left(D^2\right)^{-1}\circ \frac{M}{z}\frac{d}{dM}K
\end{equation}
which is the $M$ derivative of the field theory Feynman propagator.  Although it might seem there is a potential ordering ambiguity here, we remind the reader that $K$ involves a flat connection and so commutes with functionals of $D_\mu$.  The Ward identity then implies
\begin{equation}
M\frac{\pa}{\pa M}Z_\chi=-z\frac{N}{2}\Tr_{\Sigma\times C}\left(K^{-1}\circ D^2\circ \Delta_B\right)\,Z_\chi.
\end{equation}
Let us now lower the cutoff in $Z_\phi$.  There are two contributions
\begin{align}
M\frac{\pa}{\pa M}Z_\phi=\int[\mathscr D\phi][\mathscr D\rho]\, &\frac{i}{2z^{d-2}}\left(\underbrace{\int\phi\circ M\frac{d}{dM}K^{-1}\circ D^2\circ\phi}_{\text{Term 1}}+\underbrace{\int_{\Sigma}\phi\cdot M\frac{d}{dM}K^{-1}\cdot D_t\cdot\left.\phi\right|_{\tilde t}}_{\text{Term 2}}\right)e^{iS_\phi+iS_{source}+iS_\rho}
\end{align}
Let us massage Term 1:
\begin{align}\label{Term1}
\frac{i}{2z^{d-2}}&\int [\mathscr D\phi][\mathscr D\rho]\left(\int_C\phi\circ M\frac{d}{dM}K^{-1}\circ D^2\circ\phi\right)e^{iS_\phi+iS_{source}+iS_\rho}\nonumber\\
&=-\frac{i}{2z^{d-3}}\int [\mathscr D\phi][\mathscr D\rho]\left(\int_C\left(K^{-1}\circ D^2\circ\phi\right)\circ \Delta_B\circ K^{-1}\circ D^2\circ \phi\right.\nonumber\\
&\qquad\qquad\qquad\qquad\qquad\qquad\left.+\int_\Sigma\left.\left(\phi\cdot K^{-1}\cdot D_t\cdot \Delta_\Sigma \cdot \phi\right)\right|_{\tilde t}-\int_\Sigma\left.\left(K^{-1}\cdot D_t\cdot \phi\right)\cdot \Delta_\Sigma\cdot \phi\right|_{\tilde t}\right)e^{iS_\phi+iS_{source}+iS_\rho}
\end{align}
The boundary terms come about through integrating $D^2$ by parts.  We've defined a boundary kernel
\begin{equation}
\Delta_\Sigma:=\left.K^{-1}\cdot \frac{M}{z}\frac{d}{dM}K\right|_{\tilde t};
\end{equation}
given the form of $K$, this is a functional of only $\vec D^2$.  The last two terms of \eqref{Term1} then cancel via $D_t$ commuting through $\Delta_\Sigma$ and using $\Delta_\Sigma$ as a symmetric kernel (this is innocuous as long as $\Sigma$ has no boundary).  Terms 1 and 2 then collectively give
\begin{align}
\int[\mathscr D\phi][\mathscr D\rho]&\left(-\frac{i}{2z^{d-3}}\int_C\left(K^{-1}\circ D^2\circ \phi\right)\circ \Delta_B\circ K^{-1}\circ D^2\circ \phi-\frac{i}{2z^{d-3}}\int_\Sigma\phi\cdot \Delta_\Sigma\cdot K^{-1}\cdot D_t\cdot \phi\right)e^{iS_\phi+iS_{source}+iS_\rho}
\end{align}
Now we apply the Ward identity twice to the first term in the above expression:
\begin{align}
-\frac{i}{2z^{d-3}}\int_C\left(K^{-1}\circ D^2\circ\phi\right)\circ \Delta_B\circ K^{-1}\circ D^2\circ \phi\sim&-z\frac{N}{2}\Tr_{\Sigma\times C}\left(K^{-1}\circ D^2\circ \Delta_B\right)+z\frac{N}{2}\Tr_{\Sigma\times C}\left(\Delta_B\circ B\right)\nonumber\\
&-\frac{i}{2z^{d-3}}\int_C\phi\circ B\circ \Delta_B\circ B\circ\phi.
\end{align}
The first of these terms will be cancelled by $M\frac{\pa}{\pa M}Z_\chi$.  Finally we write
\begin{align}
M\frac{\pa}{\pa M}\mathcal Z=&\int\left[\mathscr D\phi\right]^{\varphi}\left(z\frac{N}{2}\Tr_{\Sigma\times C}\left(\Delta_B\circ B\right)-\frac{i}{2z^{d-3}}\int_C\phi\circ B\circ \Delta_B\circ B\circ \phi\right.\nonumber\\
&\qquad\qquad\qquad\qquad\left.-\frac{i}{2z^{d-3}}\int_\Sigma\varphi\cdot \Delta_\Sigma\cdot \left.K^{-1}\cdot D_t\cdot\phi\right|_{\tilde t}\right)e^{iS_\phi+iS_{source}}.
\end{align}
We want to write this expression as operators acting on $\mathcal Z$.  Most of these are straight forward, but we need to be careful with the surface term.  It might be tempting to identify $\frac{i}{2z^{d-2}}\left.K^{-1}\cdot D_t\cdot \phi\right|_{\tilde t}\sim\frac{\delta}{\delta\varphi}$ but this is not exactly correct.  There is a subtle factor of two.  To see this carefully, let us reintroduce $\rho$ to enforce the path integral boundary condition:
\begin{align}
\frac{\delta}{\delta\varphi}Z_\phi=&\int[\mathscr D\phi][\mathscr D\rho]\left(-\frac{i}{z^{d-2}}\rho\right)e^{iS_\phi+iS_{source}+iS_\rho}\nonumber\\
=&\int[\mathscr D\phi][\mathscr D\rho]\,e^{iS_\phi+iS_{source}}\left(-\frac{\delta}{\delta\phi}_\Sigma e^{iS_\rho}\right)\nonumber\\
=&\int[\mathscr D\phi][\mathscr D\rho]\,e^{iS_\rho}\frac{\delta}{\delta\phi}_\Sigma e^{iS_\phi+iS_{source}}\nonumber\\
=&\int[\mathscr D\phi][\mathscr D\rho] \left(\frac{i}{z^{d-2}}\,\left.K^{-1}\cdot D_t\cdot\phi\right|_{\tilde t}\right)e^{iS_\phi+iS_{source}+iS_\rho}
\end{align}
where we've integrated $\frac{\delta}{\delta\phi}_\Sigma$ by parts in the path integral.  Taking this into account we have
\begin{align}
M\frac{\pa}{\pa M}\mathcal Z=&z\frac{N}{2}\Tr_{\Sigma\times C}\left(\Delta_B\circ B\right)\mathcal Z-z\,\Tr_{\Sigma\times C}\left(\left(B\circ \Delta_B\circ B\right)\circ\frac{\delta}{\delta B}\right)\mathcal Z-\frac{z}{2}\int_\Sigma\varphi\cdot \Delta_\Sigma\cdot\frac{\delta}{\delta\varphi}\mathcal Z
\end{align}
Now it is simple exercise to apply this to the path integral representation of the wave functional:
\begin{equation}
M\frac{\pa}{\pa M} \Psi[z,M, B,\mathcal U; \varphi]=M\frac{\pa}{\pa M}\left(\frac{1}{\sqrt{\int[\mathscr D\varphi] \mathcal Z^*\mathcal Z}}\mathcal Z\right)
\end{equation}
In particular the $\Tr_{\Sigma\times C}\left(\Delta_B\circ B\right)$, which is a real constant will be cancelled.  Additionally the terms involving $\frac{\delta}{\delta\varphi}$ acting on the normalization can be written as a total derivative leaving a $-z\frac{N}{4}\Tr_\Sigma\left(\Delta_\Sigma\right)$ leftover.  The result of this is
\begin{align}
M&\frac{\pa}{\pa M}\Psi[z,M,B,\mathcal U;\varphi]=\left(-z\Tr_{\Sigma\times C}\left(B\circ\Delta_B\circ B\circ\frac{\delta}{\delta B}\right)-z\frac{N}{4}\Tr_\Sigma\left(\Delta_\Sigma\right)-\frac{z}{2}\int_\Sigma\varphi\cdot \Delta_\Sigma\cdot\frac{\delta}{\delta\varphi}\right)\Psi[z,M, B, \mathcal U; \varphi].
\end{align}

\providecommand{\href}[2]{#2}\begingroup\raggedright\endgroup

\end{document}